\documentclass[12pt]{article}
\usepackage[dvips]{graphicx}
\usepackage{epsf}
\def\la{\mathrel{\mathpalette\fun <}}
\def\ga{\mathrel{\mathpalette\fun >}}
\def\fun#1#2{\lower3.6pt\vbox{\baselineskip0pt\lineskip.9pt
\ialign{$\mathsurround=0pt#1\hfil##\hfil$\crcr#2\crcr\sim\crcr}}}
\title{ Progress in  Classical and Quantum Variational Principles.\footnote{Reports on Progress in Physics (2004)}}
\author{C.G.Gray \footnote{cgg@physics.uoguelph.ca}, G.Karl \footnote{gk@physics.uoguelph.ca}\\
Physics Department, University of Guelph, \\Guelph, Ontario, Canada, N1G 2W1\\
V.A.Novikov \footnote{novikov@heron.itep.ru}\\ ITEP, Moscow, Russia}
\date{}
\begin{document}

\maketitle
\newpage

\hspace*{5cm} ``...the most beautiful and important

\hspace*{5cm}  discovery of Mechanics."

\

\vspace{3mm}

\hspace*{6cm} {\it Lagrange to Maupertuis (November 1756) }

\vspace{10mm}

\begin{abstract} 
\vspace{3mm}
 We review the development and practical uses of a generalized
Maupertuis least action principle in classical mechanics, in which the action is varied
under the constraint of fixed \underline {mean} energy for the trial trajectory. The
original Maupertuis (Euler-Lagrange) principle constrains the energy at
every point along the trajectory. The generalized Maupertuis principle
is equivalent to Hamilton's principle. Reciprocal principles are also derived for both
the generalized Maupertuis and the Hamilton principles. The Reciprocal Maupertuis
Principle is the classical limit of Schr\"{o}dinger's variational principle
of wave mechanics, and is also very useful to solve practical problems in both classical and semiclassical mechanics, in complete analogy with the quantum Rayleigh-Ritz method.
Classical, semiclassical and quantum variational calculations are carried out for a number of systems, and the results are compared.
Pedagogical as well as research
problems are used as examples, which include nonconservative as well as
relativistic systems.

\end{abstract}

\newpage

\tableofcontents
\newpage

\newpage
\section{Introduction and History }
\vspace{5mm}

Variational Principles have a long and distinguished history in Physics.
Apart from global formulations of physical principles,
equivalent to local differential equations, they also are useful to
approximate problems too difficult for analytic solutions.
In the past century quantum variational calculations have been ubiquitous as an approximate  method for the ground state of many difficult systems \cite {0}.
In recent years there has been some progress in using classical variational principles
(Action Principles) to approximate the motion of classical
systems such as classical molecules. We review recent developments in this area of Classical Mechanics,
although we also overlap with Quantum Mechanics.
In particular we shall also discuss the use of the quantum variational principle for excited states,
and the connection to classical action principles.

In Classical Mechanics, variational principles are often called Least Action Principles, 
because the quantity subject to variations is traditionally the
Action. To confuse matters there are two classical Actions, corresponding
to two main Action Principles, which are called respectively,
Hamilton's Action $(S)$ and Maupertuis'
 Action $(W)$; these differ from each other (they are related by a Legendre transformation), and the notation $S,W$ is not universal
(we follow current usage but in fact some authors use the same symbols reversed \cite{0a}).
Both Actions, $S$ and $W$, have the same dimensions (i.e. energy$\times$time, or angular momentum).
 Maupertuis' Least Action Principle
is the older of the two (1744) by about a century, and as we shall
discuss, it is inconveniently formulated in all textbooks \cite{0b}. For clear statements of these
old action principles, see the textbooks by Arnold \cite{1a},
Goldstein et al \cite{1b}, and Sommerfeld \cite{1c}. We shall review these Action Principles in the next section.
Some mathematics and notation required to deal
with variational problems are summarized in Appendix I. The distinguishing feature of variational
problems is their \underline{global} character \cite{1d}.
One is searching for the \underline{function} which gives a minimum (or stationary value)
to an integral, as opposed to \underline{local}
extrema, where one is searching for the value of the variable which minimizes (
or makes stationary) a function. Of course this
dichotomy is not so sharp in practice, as the global problem is equivalent to a
differential equation (the Euler-Lagrange equation) which
is of course local. The relation has even found its way into modern
parlance \cite{Ite}: ``Think globally, act locally!''.

In Quantum Mechanics, where (usually) the aim is for approximations to the energy by
choosing a wave-function, one speaks of \underline{trial}
wave-functions, which are optimized according to a Variational Principle. In Classical
Action Principles, the equivalent notion is that of a
\underline{virtual path} or trajectory, which should be optimized in accordance with the
Action Principle. For uniformity we shall talk of
\underline{Trial Trajectories} in the classical domain, to emphasize the similarity of these ideas.

As noted above, Maupertuis' Action Principle was formulated in precise form first by Euler and Lagrange some two and a half centuries ago;
for the history, see the books by Yourgrau
and Mandelstam and by Terral \cite{2}. However, even the formulation of these eminent mathematicians remained problematic,
as emphasized by Jacobi, who stated in
his Lectures on Dynamics: `` In almost all text-books, even the best, this Principle is presented so
that it is impossible to understand'' \cite{3, 4}.
To understand at least in part the problem (see \cite{5} for a detailed discussion),
 it is sufficient to note that Maupertuis' Principle assumes
conservation of energy, whereas a well
formulated principle, like Hamilton's Principle, implies energy conservation. As we shall see,
 it is possible to
reformulate Maupertuis' Principle in a more
general form to remedy this problem and also make it more  useful in applications \cite{5,6}.
 It should be noted that Hamilton's
Principle does not suffer from any such drawbacks and is useful both conceptually
(to derive equations of motion) and as a tool for approximations.
The reformulated Maupertuis Principle has the advantage of being closely related to the
classical limit of Schr\"{o}dinger's Variational
Principle of wave mechanics, and thereby lends itself easily to semiclassical applications.

Even though we emphasize two Action Principles for Classical Mechanics, it is worth noting that there
are many other formulations and
reformulations of these principles. For a list of references on this point see  ref.\cite{5}.
We note in particular the elegant work of Percival in the 1970's on variational principles
for invariant tori  (see ref.\cite{5a} for a review).
The emphasis on Maupertuis
and Hamilton reflects the two main approaches of mechanics, one based on the Lagrangian $L$ and the other
based on the Hamiltonian $H$.

In the above and in section 2 we use the traditional terminology for the variational principles, 
i.e. ``least''  action principles, but, since Jacobi's work it
 has been recognized \cite{1a, 4} that the action is in fact \underline {stationary} in general for the true trajectories.
This means the first-order variation vanishes, and the action may be a minimum, a maximum or a saddle, depending on the
second-order variation \cite{Bru}. Hence ``stationary'' action principles would be more accurate terminology. Similar remarks apply to the
general Maupertuis and reciprocal variational principles discussed in section 2. 

In sections 2-4 we restrict ourselves to conservative holonomic systems. Various problems are solved using classical and 
semiclassical variational methods and some comparisons are made with the results from the quantum variational method for
 excited states \cite{KN}. In section 5 we discuss
nonconservative (but nondissipative) systems, and in section 6 we mention briefly particular
nonholonomic systems (which have velocity constraints), in  connection with relativistic systems.
The relations between classical and quantum variational principles (VP's) are discussed in section 7.

\section{Action Principles of Classical Mechanics}

\subsection{Statements of Action Principles}

Hamilton's Least Action Principle (HP) states that, for a true trajectory of a system, Hamilton's Action $S$ is
stationary for trajectories which  run from
the fixed initial space-time point $A\equiv {(q_A, t_A)}$  to the fixed final space-time point $B\equiv {(q_B, t_B)}$,

$$
(\delta{S})_T = 0,
\eqno(2.1)
$$
where $T = t_B - t_A$ is the duration.
Here the (Hamilton) action $S$ is the time integral of the Lagrangian $L( q, \dot {q})$ from
the initial point  to the final point  on the trial trajectory $q(t)$,

$$
S =\int\limits_{t_A}^{t_B}  \ {dt} L(q(t),\dot {q}(t)),
$$
where  $q(t)$ is the generalized coordinate, $\dot {q}(t) \equiv dq/dt$  the generalized velocity and $t$ the time. In
practice we choose $t_A = 0$ and  $t_B = T$ for convenience.
In general $q$ stands for the complete set of independent generalized coordinates $q_1, q_2, ..., q_f$, where $f$ is 
the number of degrees of freedom \cite{14a}.
\underline {In} \underline{(2.1) the constraint of fixed $T$ is indicated explicitly, but the constraint
of}$\;\;$    \underline{fixed end-positions $q_A$ and $q_B$ is left implicit.} The latter convention for the end-positions is 
also followed below in (2.2), (2.3), etc.
From the Action Principle (2.1) one can derive Lagrange's equation(s) for the trajectory,
which we shall not do in this review;
see for example \cite{1a,1b,1c}.

The reformulated (or General) Maupertuis Least Action Principle (GMP) states that, for a true trajectory Maupertuis'
Action $W$ is stationary on trajectories
with fixed end-positions $q_A$ and $q_B$ and fixed mean energy $\bar E$:

$$
 (\delta{W})_{\bar E}=0.
\eqno(2.2)
$$
Here Maupertuis' Action $W$ is given by 
$$
W =\int\limits_{q_A}^{q_B}\ p{dq} = \int\limits_{t_A}^{t_B} \ 2K{dt},
$$
where $p={\partial L}/{\partial \dot q}$ is the canonical momentum, and in general
$p dq$ stands for $p_1dq_1+ ...+ p_fdq_f$ .
 The second form for $W$ is valid for normal systems,  for
 which $L=K-V$ and $K$ is quadratic in the $\dot q$'s, where $K$ is the kinetic energy and $V$ the potential energy.
 The mean energy $\bar E$ is the time average of the Hamiltonian $H(q,p)$
 over the trial trajectory of duration $T$,
$$
\bar E =\frac {1}{T}\int\limits_{t_A}^{t_B}  \ {dt} H(q(t),p(t)),
$$
from the initial position $q_A$ to the final position $q_B$ which are common to all trial trajectories.\underline{ Note
that $\bar E$ is fixed in (2.2) but $T$ is not, the reverse of} $\;$ \underline{ the situation in (2.1).} We use the notation
$\bar E$ rather than $\bar H$ because in the early sections we
restrict \cite{note1} ourselves to dynamical variables where the Hamiltonian
 is equal to the energy $K + V$; the general case is discussed in note \cite{Cas} and later sections. 

The Reciprocal Maupertuis Principle (RMP), or principle of least (stationary) mean energy, is
\vspace{2 mm}
$$
 (\delta{\bar E})_{W}=0.
\eqno(2.3)
$$

\vspace{2mm} \noindent For stationary (i.e. steady-state or bound) motions (periodic,
quasiperiodic, chaotic) (2.3)
 is the classical limit \cite{5} of the  Schr\"{o}dinger Variational Principle
 of Quantum Mechanics: $(\delta<H>)_n = 0$, (see  also sec.7). The connection is clear intuitively: 
for bound motions, in the large quantum number
limit the expectation value of the
Hamiltonian becomes the classical mean energy and the quantum number $n$ becomes proportional
to the classical action $W$ over the  motion. This connection enables the RMP to lend itself
naturally to semiclassical applications (sec. 2.3).
For periodic and quasiperiodic motions the Reciprocal Maupertuis Principle (2.3)
is equivalent \cite{5} to Percival's principle for invariant tori \cite{5a,8a} .
The RMP is, however, more general and is valid for chaotic motions, scattering orbits, arbitrary segments of trajectories, etc.
Percival's principle has been derived by Klein and co-workers from matrix mechanics \cite{8b} in the classical limit. 
Hamilton's Principle also has a reciprocal version \cite{5} - see next section. 
For a summary of  reciprocity in variational calculus,
see Appendix I.

The textbook Euler-Lagrange version of Maupertuis' Principle differs from (2.2);
the constraint of fixed mean energy $\bar E$ is replaced by one of  fixed energy $E$,
$$
 (\delta{W})_{E}=0.
\eqno(2.4) $$

\vspace{2mm} \noindent This is not erroneous, since a true
trajectory does indeed conserve energy, but it is the source of
the inconveniences referred to by Jacobi \cite{4}. Because, as we
show in the next section, the GMP is equivalent to the HP, and
therefore to the equations of motion from which energy
conservation follows, it will become clear that energy
conservation is a consequence of the GMP (2.2), rather than an
assumption as in the original MP (2.4).

It is well known that conservation of energy is a consequence of symmetry
under time translation (Noether's theorem),
either via the Lagrangian and equations of motion \cite{8ba} ,
or from the action and Hamilton's Principle \cite{8bb}. Similarly
energy conservation can be derived directly from the GMP \cite{5}.

In the early sections we assume $L=K-V$ and that $V$ is time-independent (conservative). 
Later these restrictions are relaxed (see sections 5 and 6).

\subsection{Derivation of Action Principles}

Within Mechanics one can show that different Action Principles such as (2.1), (2.2) and (2.3)
are equivalent to each other. This we sketch below (see \cite{5} for more details).
Alternatively, one can show \cite{1a,1b,1c} from Hamilton's Principle (2.1) that the
 Lagrange equations of motion, in some coordinates are
essentially Newton's equations in the same set of coordinates. Reference \cite{2}
shows that (2.4) is equivalent to Newton's equations, quoting
an argument due to Lagrange, and in \cite{5} we derive (2.3) from similar arguments.
On a more lofty level one can derive the Action Principles from Quantum Mechanics in the classical limit.
Dirac and Feynman \cite{4'} have derived Hamilton's Principle from a path integral formulation
of Quantum Mechanics, and similarly the RMP (2.3)
can be derived from wave mechanics (see \cite{5}). Further connections between quantum
and classical variational principles are given in sec. 7.

The equivalence of Hamilton's Principle (2.1) with the General Maupertuis Principle (2.2)
is worth spelling out in a little more detail.  (The equivalence of the MP with the GMP is
discussed in Appendix II). The starting point of the derivation is the Legendre transformation
between the Lagrangian of the system and the
corresponding Hamiltonian,

$$
  H(q,p) = \sum_n p_n \dot{q}_n  - L(q,\dot{q}).
\eqno(2.5)
$$
Here $n = 1, 2, ..., f$ runs over the various degrees of freedom, and $q$ stands for $(q_1, ...,q_f)$ and
 $p$ for $(p_1, ..., p_f)$. Upon integration over time along a trial trajectory starting at $A\equiv{(q_A, 0)}$
 and ending at $B\equiv {(q_B, T)}$
 (where we could have $q_B = q_A$
 for a closed trajectory), and with the definitions
 given above we obtain

$$
\bar{E} T = W - S,
\eqno(2.6)
$$
where $T$ is the duration along the trial trajectory. Thus $W$ and $S$ are also related by
a Legendre transformation.
If we now vary the trial trajectory and the duration $T$, keeping $q_A$ and $q_B$ fixed,
and compute the variation of the
different terms in (2.6), we obtain

$$
\delta{S} + \bar{E}\delta{T} =\delta{W} - {T}\delta{\bar{E}},
\eqno(2.7)
$$

\vspace{2mm} \noindent which is the relation between the different variations. For conservative systems, near a true trajectory
Hamilton's Principle is equivalent to the vanishing of the left hand side.
In fact a vanishing  left hand side is the unconstrained form \cite{note2} of Hamilton's Principle (the UHP - see \cite{5} for proof):
 $$
\delta{S} + E\delta{T} = 0,
\eqno(2.8)
$$
where we have used the fact  $\bar{E}=E$ \underline{on} a true trajectory for 
conservative systems. Eq (2.8) has the
unconstrained form $\delta{(S+\lambda T)}= 0$,
with
$\lambda =E $ (the energy of the true trajectory) a constant Lagrange multiplier.
For fixed $T$ (i.e. $\delta{T}=0$ ), we recover the HP $(\delta{S})_T = 0$,
and for fixed $S$ (i.e. $\delta{S}=0$) we obtain the Reciprocal Hamilton Principle (RHP)

$$
(\delta{T})_S = 0.
\eqno(2.9)
$$

\vspace{2mm} \noindent The RHP (principle of least time) is
discussed in detail in ref.\cite{5}.

The right hand side of (2.7) (which therefore must also vanish near a true trajectory) gives the
unconstrained form of Maupertuis' Principle (UMP) for conservative systems:
$$
\delta{W} - {T}\delta{\bar{E}} = 0.
\eqno(2.10)
$$

\vspace{2mm} \noindent This has the unconstrained form $\delta{(W
+\lambda \bar{E})} = 0$, with $\lambda = -T$ ( the duration of the
true trajectory) a constant Lagrange multiplier. For fixed $\bar
E$ (2.10)
 gives the GMP of equation (2.2) while for fixed $W$ it gives the reciprocal,
the RMP of equation (2.3). An alternative derivation of the UMP (2.10) is given
in ref.\cite {5} with traditional variational procedures, employing the general first variation
theorem of variational calculus, which includes end-point variations \cite{Bru,Gel}.

Equation (2.7) connecting the different variations is
reminiscent of equations of thermodynamics,
where two conditions are derived from a single equation.
There are many other analogies with thermodynamics, e.g. two
actions in mechanics analogous to two free energies in
thermodynamics, with a Legendre transform relation in each case \cite{24a},
adiabatic processes  and reversible processes in both mechanics and thermodynamics, and reciprocal variational principles in both mechanics and
thermodynamics (see Appendix I). Beginning with Helmholtz \cite{2}, many of these analogies have been explored by various authors \cite{Sza}.

As we have stressed, we hold the end-positions $q_A$ and $q_B$ fixed in all the variational principles discussed above. 
It is possible to relax these constraints by generalizing 
further the UHP and UMP (see notes \cite{note2}, \cite{24a}, and \cite{24}), but we shall not need these generalizations for the applications we discuss.
 
The derivation we have sketched shows
that the four principles (2.1), (2.2), (2.3), (2.9)
are equivalent to each other. Mathematically, the Hamilton Principles
can be regarded as Legendre transformations of the Maupertuis Principles, since
the Legendre transformation relation $(2.6)$ allows us to change independent 
variables from $\bar {E}$ to $T$. 
Thus in Classical Mechanics we have a
set of four variational principles that are symmetric under Legendre and reciprocal transformations:  

$$
(\delta\bar E)_W = 0\stackrel{reciprocity}{\longleftrightarrow}(\delta W)_{\bar E} =0 \stackrel{Legendre}{\longleftrightarrow}(\delta S)_T =0\stackrel{reciprocity}{\longleftrightarrow} (\delta  T)_{S} = 0.
$$
\\
After a short interlude on action principles in semiclassical mechanics we proceed to discuss some simple examples.

\subsection{Semiclassical Mechanics from Action Principles}

  The RMP ( 2.3) is the basis of a very simple semiclassical quantization method. As we
shall see, the solution of (2.3) yields, for the true trajectory or approximations to it,
 $\bar E$ as a function of the action $W$, $\bar E(W)$.
For bound motions we then assume the standard Einstein-Brillouin-Keller
(EBK) \cite{9a} quantization rule
for the action over a cycle
$$
 W = (n+\alpha)h,
\eqno(2.11)
$$
where $n=0,1,2,...$ ,  $\alpha$ is essentially the Morse-Maslov index ( e.g. $\alpha=1/2$ for a 
harmonic oscillator),
and $h \equiv 2 \pi \hbar$ is Planck's constant. (EBK (or torus) quantization is the generalization of
Bohr-Sommerfeld-Wilson quantization
from separable to arbitrary integrable systems). Thus we
have expressed $\bar E$ as a function of the quantum number $n$, $\bar E= \bar E_n$.
For multidimensional systems, (2.11) is applied to each action $W_i$ in  $\bar E(W_1, W_2,...)$,
 so that we get $\bar E_{n1,n2,..}$ as a function of the quantum numbers $n_i$.
This method has been found to be reasonably accurate even for nonintegrable systems, where strictly
speaking the good actions $W_i$ do not exist. The UMP (2.10) can also be used to obtain semiclassical expressions for  $\bar E_{n1,n2,..}$. Examples are given later in sections 4.1 - 4.3.

\section{\bf Practical Use of Variational Principles. Pedagogical Examples}

Here we discuss how to use variational principles to solve practical problems. We use the so-called direct
method \cite{Gel} of the calculus of variations (e.g. Rayleigh-Ritz), which operates directly with the variational
principle and makes no use of the associated Euler-Lagrange differential equation. Historically \cite{note6}, the method originates with Euler,
 Hamilton, Rayleigh, Ritz and others (see \cite{5, Gel} for some
 early references).We assume a trial solution, which contains one or more  adjustable parameters $a_i$.
We then use the variational principle to optimize the choice of the $a_i$. Generally speaking, the more parameters
$a_i$, the better the solution. The direct method is well known in quantum mechanics, but is also very useful in classical mechanics, as we shall see. We give a few
simple examples in this section, mainly for bound motions; further simple examples, including for
scattering problems, are given in refs. \cite{5,6} and in the references therein .

The simplest example is the free motion of a particle of unit mass on the
line segment $x = 0$ to $x = 1$.  With the traditional Maupertuis principle
(2.4) once we fix the energy $E = v^2 /2$ there is essentially nothing left to vary. With
the GMP (2.2) or its reciprocal (2.3) we can choose a variety of trial trajectories. For example we
can take velocity $v_1$ on the sub-segment $0 < x < 1/2$
 and velocity $v_2$ on the sub-segment $1/2 < x < 1$. The action $W$ is
$(v_1 + v_2)/2$ and the mean energy $\bar E$ is $v_1v_2/2$, where we can vary
$v_1$ to change $W$ while keeping the mean energy fixed. We then find that
$v_1 = v_2$, a simple example of energy conservation for this particular trial trajectory. By iteration we can now see
that energy is conserved for the whole trajectory on the segment $(0,1)$, as a consequence
of the General Maupertuis Principle (2.2).

\subsection{The Quartic Oscillator}

An instructive example is the one-dimensional (1D) quartic oscillator, with Hamiltonian

$$
H= \frac{p^2}{2m} + \frac{1}{4} C x^4.
\eqno(3.1)
$$
Using a harmonic oscillator trial trajectory with
$$
x(t) = A \sin{\omega}t,  \;\;\; p(t) =  m A {\omega}\cos{\omega}t
\eqno(3.2)
$$
we find for a complete cycle with period $T= {2\pi}/{\omega}$

$$
\bar E  = \frac{\omega}{4\pi}W + C \frac{3W^2}{32{\pi}^2 m^2 {\omega}^2},
\eqno(3.3)
$$

$$
W = {\pi\omega}m A^2.
$$

From (3.3), with the use of the RMP (2.3) we can obtain the ``best" frequency
$\omega_0$ solving $({\partial \bar E}/{\partial \omega})_W = 0$,  which can be substituted into (3.3)
to obtain $\bar E$
$$
\omega_0 = [\frac {3 C W}{4\pi m^2}]^{1/3}, \;\;\;  \bar{E} =\frac{1}{2}(C/{m^2})^{1/3} ({3 W}/{4 \pi})^{4/3}
\eqno(3.4)
$$
and
the period $T$ as a function of  mean energy:
$$
T(\bar{E}) = 2\pi [ {m^2}/(2 C \bar{E})]^{1/4}.
\eqno(3.5)
$$
This variational estimate for the period is in error by $0.75\%$ when compared with the exact result,
 as noted in reference \cite{6}. Systematic improvements can be obtained by including terms
 $B \sin{3\omega}t$, $C \sin{5\omega}t$, etc., in the trial trajectory $x(t)$; see \cite{6} for a related example.

We have used the RMP $(\delta \bar E) _W=0$ here since (3.3) made it very
convenient to do so. The GMP $(\delta W)_{\bar E} =0$ and the HP
are also viable for this problem, but the original MP
$(\delta W)_E =0$ is not, since the constraint of fixed $E$ leaves
essentially no freedom for variation for 1D problems \cite {21a}. One
can get around this by relaxing the constraint  via a Lagrange
multiplier, but this is then equivalent to the GMP (see App. II) or the 
UMP. In 2D etc., the constraint of fixed $E$ does allow some freedom
for variation, but the constraint is very cumbersome and 
the GMP is always much more convenient.

The use of the action principle (2.3) to obtain the mean energy as a function of the action (3.4) or
the period as a function of mean energy (3.5), with harmonic oscillator trial
trajectories, is very similar to the use of the variational principle in quantum mechanics
to estimate the energies of states of the quartic oscillator with  harmonic oscillator trial
wave-functions. We show this explicitly for the eigenstates of the
Hamiltonian (3.1). In both cases an optimum frequency for the trial
harmonic case is sought and the final errors are also similar.
We start by computing the quantum expectation value of (3.1) using a harmonic oscillator
state with trial
frequency $\omega$,
$$
<n|H|n> =\frac{1}{2}(n+ \frac{1}{2})\hbar\omega + \frac{3C}{8}(\frac
{\hbar}{m\omega})^2 (n^2+n+\frac{1}{2}),
\eqno(3.6)
$$
and find the optimum frequency $\omega_0$ by demanding the vanishing of the derivative of $<H>$ with respect to the
frequency. We substitute this frequency in (3.6) to estimate the energy of excited states

$$
\bar{E}_n = \frac{3}{4}(\hbar)^{4/3}(\frac {3 C}{2 m^2})^{1/3}(n^2+n+\frac{1}{2})^{1/3} (n+ \frac{1}{2})^{2/3}.
\eqno(3.7)
$$

\vspace{2mm} \noindent Taking the asymptotic limit of (3.7) for $n
\gg 1$ and replacing $2 \pi \hbar n$ by $W$ we obtain the second
equation (3.4). This is not as trivial as it seems, since (3.7)
was obtained from the quantum variational principle (for an
arbitrary state \cite{KN}), while (3.4) was obtained from a
classical action principle.

This example illustrates neatly that the two variational principles are related: the Reciprocal Maupertuis Principle
is the classical limit of the Schr\"{o}dinger Variational Principle (see sec.7 for further discussion).
The example illustrates also that the classical action principle can be used as a method
of approximation, and semiclassically to estimate $E_n$ as described in sec.2.3.
According to (2.11), we quantize $\bar {E}(W)$ in (3.4) semiclassically by the replacement 
$W\to (n+\frac {1}{2})h$; as just discussed,
this leads to agreement asymptotically ( $n \to \infty$) with the quantum result (3.7).
Reference \cite{KN} discusses the similar relation between quantum and classical variations for a different one-dimensional system,
with a linear potential. The application of classical action principles to other simple systems is discussed in reference \cite{6}.
The application of quantum variations for excited states has a long history: the earliest reference is McWeeny and Coulson,
who applied it to the quartic oscillator described above \cite{8c} more than fifty years ago (for later references see \cite{KN}).

\subsection{The Spherical Pendulum. Precession of Elliptical Orbits }

For a pendulum with two degrees of freedom ($\theta, \phi$), there
is an additional term in the kinetic energy compared to that of a
plane pendulum, i.e.
$$
K = \frac{1}{2}mL^2(\dot\theta^2 +\sin^2 \theta\dot\phi^2) \; ,
\;\; V = mgL(1-\cos\theta) \;\; ,
\eqno(3.8)
$$
where $\theta$ is the polar angle (measured from the downward vertical axis)
 and $\phi$ is the azimuthal angle, $m$ is the bob mass,  $L$ the length, and $g$ the gravitational acceleration.

Using the axial symmetry around the $z$ (vertical) axis, we
introduce the cylindrical coordinate $\rho = (x^2 + y^2)^{1/2} =
L\sin\theta$, and the angular momentum component $l_z =
m\rho^2\dot\phi$, which is a constant of the motion. We also expand
$\cos\theta$, keeping up to quartic terms, which are $O(L^{-2})$.
This gives
$$
K=\frac{1}{2}m\dot\rho^2(1+\frac{\rho^2}{L^2}) +\frac{l_z^2}{2m\rho^2} \; ,
\;\; V=\frac{1}{2} m\omega_0^2 \rho^2(1+\frac{1}{4}\frac{\rho^2}{L^2}) \;\; ,
\eqno(3.9)
$$
where $\omega_0^2 = g/L$. With two degrees of freedom and two constants of the motion 
($l_z$ and the energy), the system is integrable.

If the quartic terms are neglected in (3.9) by assuming
$L\to\infty$, the $(x,y)$ or $(\rho,\phi)$ motion is that of a 2D
isotropic harmonic oscillator with frequency $\omega_0$. In
particular, an elliptical orbit, with semiaxes $a$ and $b$, is
fixed in space, and the motion is periodic. If the quartic terms are not negligible, the
oscillator frequency shifts from $\omega_0$ to a lower value
$\omega$, as in the plane pendulum case, where
$$
\omega = \omega_0 \left(1-\frac{a^2 +b^2}{16L^2}\right) \;\; ,
\eqno(3.10)
$$
and the orbit precesses in the prograde sense with frequency
$\Omega$, where
$$
\Omega = \frac{3}{8} \frac{ab}{L^2}\omega_0 \;\; .
\eqno(3.11)
$$
The motion is now in general quasiperiodic. The precession is one of the major sources of error to be
eliminated  in the design of Foucault pendula \cite{13a}. (The  Foucault pendulum
is discussed in Sec. 5.2.2.) Because
the precession rate is $O(L^{-2})$, this precession is particularly
relevant for recently developed short Foucault  pendula, which
are less than 1 m in length. Equations (3.10) and (3.11) have
been derived by perturbation theory \cite{13b}.

To derive (3.10) and (3.11) variationally \cite{6}, we first choose a trial
trajectory; we take an ellipse which is closed in the coordinate
system $x^\prime, y^\prime$ which rotates with respect to the
fixed $x,y$ system with  the (unknown) precession frequency $\Omega$. In the
$x^\prime, y^\prime$ system we therefore have $x^\prime =
a\cos\omega t$ and $y^\prime = b\sin\omega t$, where $\omega$ may
differ from $\omega_0$. Because the $x^\prime, y^\prime$ and $x,
y$ systems differ by a rotation, we have $\rho^2 = x^2 + y^2 =
x^{\prime^2} + y^{\prime^2}$, and therefore our trial trajectory
in the fixed coordinate system is
$$
\rho^2 = a^2\cos^2 \omega t + b^2 \sin^2 \omega t \;\; .
\eqno(3.12)
$$

To apply the Maupertuis principles, we calculate the action $W$
and mean energy $\bar E$ over one period $T=2\pi/\omega$ of the
rotating ellipse,
$$
W = T \langle 2K \rangle,\;\;\;   \bar E = \langle K +V\rangle \;\; ,
\eqno(3.13)
$$
where $\langle ...\rangle$ denotes a time average over one period,
and $K$ and $V$ are given by (3.9). The averages of the terms in
(3.9) are easily calculated using (3.12), which gives
$$
\langle\rho^2\rangle = \frac{1}{2}(a^2 +b^2) \; , \;\;
\langle\dot{\rho}^2\rangle = \frac{\omega^2}{2}(a-b)^2 \; , \;\;
\langle\frac{1}{\rho^2}\rangle = \frac{1}{ab} \; ,
$$
$$
~~~
\eqno(3.14)
$$
$$
\langle\rho^2 \dot{\rho}^2\rangle = \frac{\omega^2}{8}(a^2 -b^2)^2 \; ,
\;\; \langle\rho^4\rangle = \frac{3}{8}(a^2 -b^2)^2 +a^2 b^2 \;\; .
$$
\vspace{7mm}

The precession frequency $\Omega$ can be related to the angular
momentum using $\omega +\Omega \equiv \langle\dot\phi\rangle = \langle
l_z/m\rho^2\rangle$, and $\langle 1/\rho^2\rangle = 1/ab$ from
(3.14), giving
$$
l_z = m ab (\omega +\Omega) =
m ab \omega\left(1+\frac{\Omega}{\omega}\right) \;\; .
\eqno(3.15)
$$

\begin{figure}[h]
\begin{center}
\includegraphics[width=5cm]
{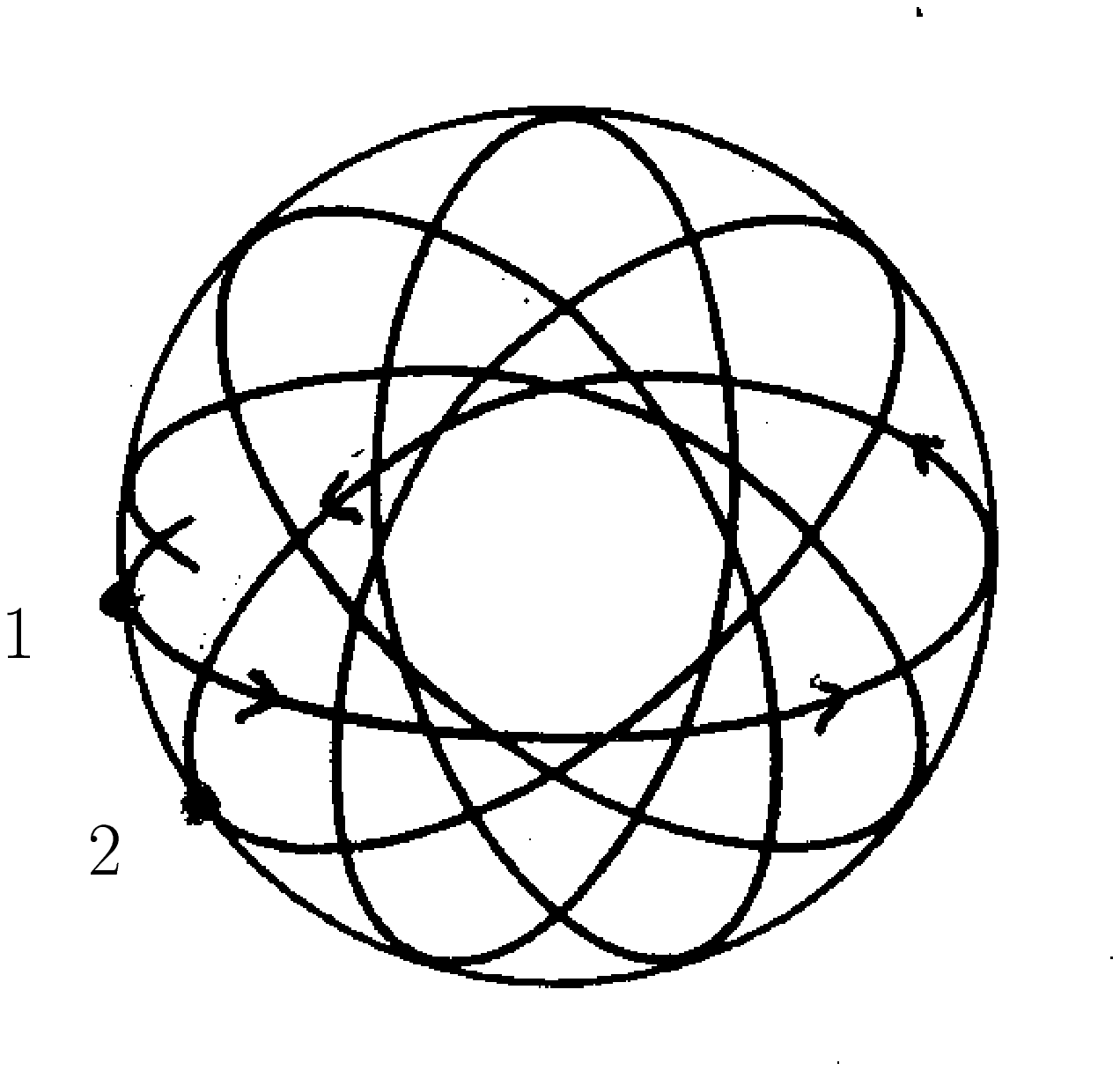}
\end{center}
\caption{The precessing orbit.  We choose positions 1 and 2 as the fixed end positions to be used in the action principle.
(figure adapted from ref.\cite{1a})}
\end{figure}

As initial and final positions 1 and 2 (see Fig.1), we take two successive aphelia
at $x^\prime =a$, i.e. $(\rho_1, \phi_1) = (a, 0)$ and $(\rho_2,
\phi_2) = (a, \phi_2)$, where $\phi_2 = (\omega +\Omega)T =
2\pi(1+\Omega/\omega)$.The positions 1 and 2 and hence $a$ and
$\Omega/\omega$ are thus fixed, which leaves $b$ and $\omega$
available as variational parameters. Note that  (3.15) then
implies that $l_z$ is varied with the trial trajectory (although
it is a constant of the motion on the individual trial
trajectories). We find it easiest to use the unconstrained version
of the Maupertuis principles [i.e. the UMP (2.10)], $\delta\bar E =
T^{-1}\delta W$, where $T = 2\pi/\omega$ is the time to reach
position 2 from position 1. Thus we use
$$
\frac{\partial}{\partial\omega}\bar E = (\frac{\omega}{2\pi})
\frac{\partial}{\partial\omega} W \; , \;\;
\frac{\partial}{\partial b}\bar E = (\frac{\omega}{2\pi})
\frac{\partial}{\partial b} W \; ,
\eqno(3.16)
$$
where, choosing units such that $m=1$, $L=1$, $\omega_0 =1$,
we have from (3.13) to (3.15)
$$
\bar E = \frac{\omega^2}{4}(a-b)^2 +\frac{\omega^2}{16}
(a^2 -b^2)^2 +\frac{\omega^2}{2}(1+\frac{\Omega}{\omega})^2 ab
$$
$$
+\frac{1}{4}(a^2 +b^2) +\frac{3}{64}(a^2 -b^2)^2 +
\frac{1}{8}a^2 b^2 \; ,
\eqno(3.17)
$$
and
$$
\frac{W}{4\pi} =\frac{\omega}{4}(a-b)^2 +\frac{\omega}{16}(a^2 -b^2)^2
+\frac{\omega}{2}(1+\frac{\Omega}{\omega})^2 ab \;\; .
\eqno(3.18)
$$
\vspace {5mm}

The first of the variational equations (3.16) is satisfied
identically by (3.17) and (3.18), and gives no useful information.
From the second, setting $\omega = \omega_0 +\Delta = 1+\Delta$, where $\Delta$ is the frequency shift,
and retaining terms only to second order in $a$ and $b$, i.e.,
$O(L^{-2})$ in general units, we get the condition
$$
\frac{a}{b}\Omega +\Delta = \frac{5}{16}a^2 -\frac{1}{16}b^2 \;\; .
\eqno(3.19)
$$

When $a=0$, (3.19) gives $\Delta = -b^2/16$ as it should [recall
the plane pendulum result \cite{5, 6}]. By symmetry, $\Delta$ must
reduce to $-a^2/16$ when $b=0$. This requires that $\Omega =
(3/8)ab$ in (3.19), and from this result we then get $\Delta =
-(a^2 +b^2)/16$ from (3.19). When we restore the dimensional
quantities $\omega_0$ and $L$, these results agree with (3.10) and
(3.11), so that with a harmonic oscillator trial trajectory, the
variational method gives results correct to $O(L^{-2})$.

Hamilton's Principle can also be applied to this problem. Here, we
need the Hamilton action $S = \int^T_0 Ldt = T\langle K-V\rangle$,
where $T$ (and hence $\omega$) is now fixed, in addition to the
positions 1 and 2 (and hence $a$ and $\Omega/\omega$). This leaves
$b$ available as a variational parameter, and setting $\partial
S/\partial b =0$ leads to the same results as above.

\section{More Complicated  Examples. Research Problems}

More complicated examples are easy to find, and are especially interesting in multidimensional 
systems which exhibit chaotic motion. Such systems are nonintegrable, and have fewer constants
of the motion than the number of degrees of freedom. We illustrate the use of classical and semiclassical variational principles together with the application of quantum 
variational principles to excited states.

\subsection{The $ x^2 y^2 $ oscillator}

The  2D $x^2y^2$ oscillator is a simple example of this class,
which has been studied in the literature for some time (see, e.g. \cite{9,Whe}).
There is only one constant of the motion, the energy.
Classically, most trajectories in this system are chaotic \cite{10}.
Fig.2 shows the contours of the $x^2y^2$ potential, and Fig.3 shows a typical (chaotic) classical trajectory.

\begin{figure}[h]
\begin{center}
\includegraphics
{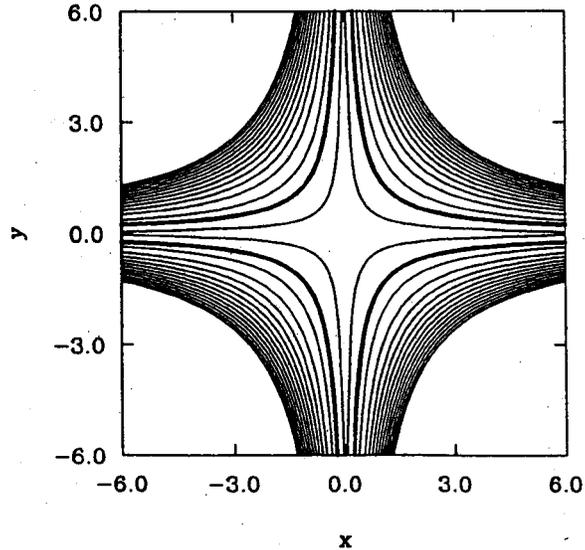}
\end{center}
\caption{  Contours of the potential $V(x,y)=x^2y^2/2$. The darkened contour
corresponds to $V(x,y)=1$. (from ref.\cite{9}) }
\label{fig:2}
\end{figure}

\begin{figure}[h]
\begin{center}
\includegraphics
{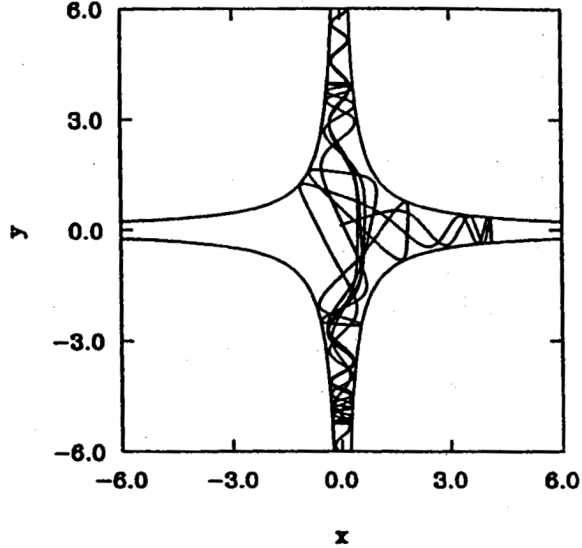}
\end{center}
\caption{  A typical trajectory of the Hamiltonian (4.1). The motion starts near the origin, then proceeds into
the right-hand arm of the potential along the positive $x$-axis. The trajectory eventually reverses
its  $x$-direction and returns to the vicinity of the origin before entering another arm
of the potential. (from ref.\cite{9}) }
\label{fig:3}
\end{figure}

Refs.\cite{5,6} give classical and semiclassical discussions of the motion using the RMP.
Here we discuss the quantum states using variational methods for the
excited states, as done above for a simpler example. Comparison with the semiclassical results
is given later.
The Hamiltonian is

$$
H = K_x + K_y + V(x,y) = \frac{p_x^2 +p_y^2}{2} +\frac{x^2 y^2}{2}.
\eqno(4.1)
$$
We choose units where $m=1$ and $C=1$ , where  ${C x^2 y^2}/{2}$ is the potential. Below we also choose
$\hbar=1$.
We take as an approximation a separable trial wave-function $\psi_{n_x,n_y} (x, y)$
$$
\psi_{n_x,n_y} (x,y) = \psi_{n_x} (\omega_{x};x)\psi_{n_y} (\omega_{y}; y) ,
\eqno(4.2)
$$
where the wave-function  $\psi_{n_x} (\omega_{x}; x)$ 
is an eigenstate of a harmonic oscillator with frequency $\omega_x$:
$$
 \frac{p_x^2 +\omega_x^2 x^2}{2}\psi_{n_x}(\omega_x; x) = (n_x+\frac{1}{2})
\omega_x \psi_{n_x}(\omega_{x}; x).
\eqno(4.3)
$$

The frequencies $\omega_x$ and $\omega_y$  are variational parameters, which will differ from state to state,
 and as a result the set (4.2) is nonorthogonal.
The computation of matrix elements and expectation values with the wave functions (4.3) is rather easy,

$$ <n|K|n>  = \frac{1}{2}(n+\frac{1}{2})\omega,  \; \; \;\; <n|x^2|n>
= \frac{2n+1} {2\omega}, \eqno(4.4) $$
and therefore the
expectation value of the Hamiltonian (4.1) with the wave-functions
(4.2) is $$ <H> = <n_x,n_y| H |n_x,n_y> =
\frac{1}{2}[(n_x+\frac{1}{2})\omega_x + (n_y+\frac{1}{2})\omega_y
+\frac{(n_x+\frac{1}{2})(n_y+\frac{1}{2})}{\omega_x\omega_y}] .
\eqno(4.5) $$ We then find the best frequencies $\omega_x$ and
$\omega_y$  by setting to zero the partial derivatives of $<H>$
with respect to the two $\omega$'s to obtain the frequency
conditions

$$
\omega_x^2\omega_y = (n_y+\frac{1}{2}) \; , \;\;\;\omega_x\omega_y^2 = (n_x+\frac{1}{2}),
\eqno(4.6)
$$
from which we obtain

$$
(n_x+\frac{1}{2})\omega_x = (n_y+\frac{1}{2})\omega_y.
\eqno(4.7)
$$
Eqs. (4.7) and (4.3) show that at the variational minimum the ratio of the two frequencies is rational and the mean energy of the
$x$-oscillator is the same as the mean energy of the $y$-oscillator. It follows immediately that the mean kinetic
energies of the two oscillators are also equal to each other at the variational minimum: $<K_x> = <K_y>$.
From (4.6) we also find
$$
\omega_x\omega_y = [(n_x+\frac{1}{2})(n_y+\frac{1}{2})]^{1/3},
\eqno(4.8)
$$
which gives the separate frequencies

$$
\omega_x = (n_x+\frac{1}{2})^{-1/3}(n_y+\frac{1}{2})^{2/3}, \;\;\;
\omega_y = (n_x+\frac{1}{2})^{2/3}(n_y+\frac{1}{2})^{-1/3},
\eqno(4.9)
$$
and the variational estimate for the energy:
$$
E_{n_x,n_y} = <n_x,n_y| H |n_x,n_y> = \frac{3}{2} [(n_x+\frac{1}{2})(n_y+\frac{1}{2})]^{2/3}.
\eqno(4.10)
$$
Therefore the variational states are labelled by the two integers $n_x$ and
$n_y$ and the energy at high quantum numbers increases as the two thirds
power of their product. 

The estimate (4.10) can also be obtained 
semiclassically by using the RMP to find 
$\bar{E}(W_x, W_y)$, and then quantizing the actions
$W_i$ with the EBK relation (2.11); see ref.\cite{6} for a detailed discussion. It is to be noted that this method yields 
the quite reasonable first approximation (4.10) for $E_{n_x,n_y}$ despite the fact that strictly speaking
 the good actions $W_x$ and $W_y$ do not exist for the $x^2y^2$ oscillator. In essence, one is approximating a chaotic trajectory by a quasiperiodic one, and the RMP picks out the best such trial trajectory \cite{35a}. 

The estimate (4.10) differs only in
the leading coefficient , 3/2 vs. 1.405, from a semiclassical adiabatic formula
(SCA) obtained by Martens et al \cite{9}.
The difference between the two estimates is about $ 6.5\%$. As noted by Martens et al \cite{9} formula (4.10)
gives rise to degeneracies.
For example $(n_x,n_y)$ = (7,0), (0,7), (2,1), (1,2) give the same energies - a four-fold degeneracy.
However, numerical estimates of the first 50 energy levels, obtained in \cite{9},  do not show such high degeneracies.
Also, because the symmetry group of the Hamiltonian
$C_{4v}$ (a fourfold axis of rotation, and four reflection planes)
has only one- and two-dimensional irreducible representations  \cite{Tin}, we expect at most two-fold degeneracy.
We shall see that the basis (4.2) permits an evaluation of the splitting of these spurious degeneracies.
The pattern of degeneracies is easy to understand if the bracket is rewritten 
$$
[(2n_x+1)(2n_y+1)/4] = N/4,
$$
where the number $N$ (the``principal" quantum number), is an odd integer. The degeneracy pattern is determined
by the decomposition of $N$ into prime factors. For example
$(n_x,n_y)$=(2,1) corresponds to $N$ =15 which can be factored as $15 \cdot1$, $1 \cdot15$, $3 \cdot 5$, $5 \cdot3$
and these products correspond to (7,0) etc. If $N$ is a prime number there are only two states,
corresponding to $N \cdot 1$ and $1 \cdot N$, while every nontrivial factorization
of a nonprime $N$ allows further states. The integer $N$ belongs to one of two possible classes: $N= 4k+1$ or $N=4k-1$. For the
case $4k-1$ (e.g. $N$ =15) one of $n_x,n_y$ is even and the other is odd. This class of states EO and OE (of \cite{9})
corresponds to the two dimensional irreducible representation of  $C_{4v}$. For
$ N=4k+1$ ( e.g. $N$=21) the two quantum numbers $(n_x, n_y)$ are both even or both odd
(e.g.  $(n_x, n_y)$ = (10,0) or (0,10) or (1,3) or (3,1) for $N$=21, which is therefore four-fold degenerate).
This degeneracy is split into states of symmetry EEE, EEO, OOE and OOO, in the notation of Martens et al \cite{9}
for the one dimensional irreducible representations.
In this notation the first two letters give respectively the parity ( Even or Odd) of $n_x$ and $n_y$
(or equivalently the behaviour of the wave-function under the reflections ($ x \to -x $, $y \to -y$), while the third
letter stands for the behavior under the interchange of $x$ with $y$. We discuss later the splitting of degeneracies
resulting from the expression (4.10). At this point we wish only to note that
the degeneracy grows very slowly with $N$, like  lnln$N$,
and even for $N$ quite large where quasiclassical behaviour is expected (say $N\sim 105$ )
the typical number of degenerate
states is quite modest, 4 - 6.

 The main points of discussion arising from the result (4.10) are
the splitting of the degenerate
levels, perturbative corrections to the variational energies , and the density of states implied by this formula.
For the first fifty low lying states one has the accurate numerical estimates of Martens et al \cite{9}
with which to compare.

\subsubsection{Perturbative Shift of Variational Results}

We can improve the variational result (4.10) if we rewrite the Hamiltonian $H$ (4.1) in the form
$$
H = H_0 + V = K +  \frac{\omega_x^2 x^2 + \omega_y^2 y^2}{2}
+ \frac{x^2 y^2- \omega_x^2 x^2 - \omega_y^2 y^2}{2}.
\eqno(4.11)
$$
Here, the perturbation $V$ is the last term above, and its effect  is already included in first order, in the variational estimate (4.10).
Therefore the first correction comes in second order in $V$. We must also remember that that the frequencies
 $(\omega_x,\omega_y)$ are appropriate to a fixed pair of quantum numbers
 $(n_x,n_y)$ , and change if we go to a different pair. The simplest way to do perturbation
 theory in these circumstances is to estimate corrections to each state  $(n_x,n_y)$ by considering a set of orthogonal intermediate states
$(n_x^{'},n_y^{'})$  all with the same frequency $(\omega_x,\omega_y)$  appropriate to the state $(n_x,n_y)$  which we consider.
Therefore we take the second order correction

$$
\Delta E_{n_x,n_y}^{(2)} = -\sum \frac{|<n_x^{'},n_y^{'}| V |n_x,n_y>|^2}{[(n_x^{'} - n_x)\omega_x + (n_y^{'} - n_y)\omega_y]}.
\eqno(4.12)
$$

\vspace{2mm} \noindent
Due to the form of $V$ (quadratic in the coordinates $x,y$), we must have $n_x^{'} = n_x$
or $n_x^{'} = n_x + 2$ or $n_x^{'} = n_x - 2$ for nonzero contributions and similarly for $n_y^{'}$. It is an interesting
property of the variational solution that only intermediate states with both $n_x^{'}$ different from $n_x$ and
$n_y^{'}$ different from $n_y$ give nonvanishing contributions, because of the frequency conditions. This can be seen
from an evaluation of matrix elements with only one quantum number different, for example $n_x^{'} \not= n_x$, but
$n_y^{'} = n_y$. Then one finds

$$
<n_x^{'},n_y| V |n_x,n_y> = <n_x^{'}|\frac{x^2}{2}|n_x>[<n_y|y^2|n_y> - \omega_x^2],
\eqno(4.12^{'})
$$
and the bracket on the right hand side vanishes because of (4.6). Only the perturbation $x^2y^2/2$ contributes
to nonvanishing matrix elements. There are only four intermediate states which contribute: $(+,+),(+,-),(-,+),(-,-)$, where
$ (+,-) = (n_x+2,n_y- 2)$, etc. Therefore (4.12) becomes after a little computation
$$
\Delta E_{n_x,n_y}^{(2)} = -\frac{1}{24} E_{n_x,n_y}^{(0)},
\eqno(4.13)
$$
and
$$
E_{n_x,n_y}^{(2)} = \frac {23}{16}[(n_x + 1/2)(n_y + 1/2)]^{2/3},
\eqno(4.14)
$$
 where $\Delta E^{(2)}$ denotes a second-order shift and $ E^{(2)}$ denotes the energy up to the second-order.

Therefore, in second-order perturbation theory, the coefficient $3/2$ is
diminished by about $4\%$ to 1.437, closer to the adiabatic result 1.405 of Martens et al \cite{9} .
 The remaining difference is about $2.5\%$.
The degeneracies of (4.14) are still the same as those of (4.10).
If we apply perturbation theory to the ground state $(0,0)$ we find

$$
E_{00} = E_{00}^{(0)} + \Delta E_{00}^{(2)} + \Delta E_{00}^{(3)} + \Delta E_{00}^{(4)}
      =0.5953 -0.0248 + 0.0496-0.1465 +...
\eqno(4.15)
$$
The sequence (4.15) indicates that the perturbation expansion does not converge, and it is
a good idea to stop at second order.
The accurate numerical result for $E_{00}$ (from \cite{9}) is 0.5541, which shows that
stopping at second order gives an error of
about $3\%$ for the ground state.

 Some examples of the estimates $E^{(0)}, E^{(2)}$, and $E_{SCA}$ are given in Table 1,
and compared with numerical estimates from \cite{9},
for a set of states which are also relevant to the next section.

\vspace{15mm}
\begin{center}

{\bf Table 1}

\vspace{3mm}

\begin{tabular}{|c|c|c|c|c|c|c|}
\hline

$N$ & $(n_x,n_y)$ & $E_{exact}$ & $E_{var}^{(0)}$ & $E^{(2)}$ & $E_{split}$ & $ E_{SCA}$\\
\hline
 1  &  (0,0)  &   0.5541   &    0.595   &   0.570  &       $ na$  &   0.5577 \\

 5 & (2,0)  & 1.7575   &   1.7406   &   1.6681    &   1.7471   &     1.6308 \\

 9   &  (4,0)  &  2.4920  &   2.5756    &  2.4683   &     2.5377   & 2.4132 \\

13 & (6,0)  & 3.1180   &  3.2911    & 3.1540   &  3.2171    &    3.0836 \\

17 & (8,0)    & 3.6909   &  3.9357    & 3.7717  &   3.8309  &  3.6874 \\

21 & (10,0) & 4.2504    &  4.5310  &   4.3422   &  4.3973   &  4.2453 \\

25  & (12,0) &   5.0490   & 5.0895    &  4.8775   &  4.9297  &   4.7685 \\

29  & (14,0)  &  5.3327  &   5.6189  &   5.3847  &  5.4347  & 5.2645 \\
\hline
\end{tabular}

\end{center}

{ EEE states at low excitation. The exact values and $E_{SCA}$ are taken from \cite{9};
the variational estimates and
second order results correspond to formulae (4.10) and (4.14), while
the $E_{split}$ are obtained following section 4.1.2}

\subsubsection{Splitting of Degenerate States}

The basis (4.2) is also quite convenient to obtain the splitting of degenerate states. We shall deal with the simplest
example with a degeneracy: the pair of states $(2,0)$ and $(0,2)$. Both these states have the same energy
$(3/2)(5/4)^{2/3}$ in the variational approximation. The wave-functions of the two states are nonorthogonal, because
the frequencies in the $x$ and $y$ oscillators differ. We define the overlap integral $\Delta$

$$
        \Delta = <20|02> = |\int{dx}\psi_{2}(\omega_2;x)\psi_0(\omega_0;x)|^{2} =\frac{4}{27}(\frac{5}{4})^{1/2},
\eqno(4.16)
$$
where the two wave functions have different frequencies
( $\omega_2 =(1/10)^{1/3}$ and $\omega_0=(25/2)^{1/3}$) as denoted.
The splitting of two (or more) nonorthogonal
degenerate states is well known; a simple example is the Heitler-London treatment of the hydrogen molecule.
One has to take the sum or the difference of the two nonorthogonal states , e.g. $ \Psi_{+} = N_{+}[\psi_{02} + \psi_{20}]$,
and similarly for $\Psi_{-}$. The eigenvalues have the form
$$
E_{+} =(H_{AA} + H_{AB})/(1+\Delta),
$$
  and   $$ E_{-} = ( H_{AA} - H_{AB})/(1-\Delta),
\eqno(4.17)
$$
where here  $|A>\equiv|2,0>$ and $|B>\equiv|0,2>$.

Thus in our case we need the off-diagonal matrix element of the Hamiltonian between the two states $(0,2)$ and $(2,0)$.
This can be obtained from the overlap integral (4.16)
$$
<20| K_{x} + \frac{\omega_2^2x^2}{2} |02> = \frac{5}{2}\omega_2\Delta, \;\;
<20| K_x + \frac{\omega_0^2 x^2}{2}|02> = \frac{1}{2} \omega_0 \Delta.
\eqno(4.18)
$$
 If we subtract the two equations $(4.18)$ we obtain
$$
<20| (\omega_2^2 - \omega_0^2) \frac{x^2}{2}|02> = ( 5\omega_2-\omega_0)\Delta.
\eqno(4.19)
$$

The right hand side of $(4.19)$ vanishes because of the frequency condition $(4.7)$ and therefore the matrix element
of $x^2$ vanishes as well between the two variational states. The same is true for matrix elements of $y^2$, and $x^2 y^2$.
Therefore only the kinetic energy terms contribute to the off-diagonal matrix elements of the Hamiltonian . Using
equations $(4.18)$ we find
$$
<20| K_x |02>  = 12\frac{\omega_2^2\omega_0\Delta}{\omega_0^2 - \omega_2^2} =
\frac{1}{2}\omega_0\Delta.
\eqno(4.20)
$$

The same value is obtained for the matrix element of $K_y$ , and therefore we find
$$
<20|H|02> = \omega_0\Delta.
\eqno(4.21)
$$
 In our case
$\omega_0 = (4/3)E_{02}$, and therefore we have for the two split eigenvalues
$$
E_{+} = E_{02}\frac{1 + (4/3)\Delta}{1 + \Delta} = 1.8230;\;\;\;
E_{-} = E_{02}\frac{1 - (4/3)\Delta}{1 - \Delta} = 1.6254.
\eqno(4.22)
$$
Hence the two degenerate states at $E_{02}=1.7406$ split into an EEE state at
1.823 and an EEO state at 1.625.
We identify $\Psi_{+}$ as an  EEE state because $\Psi_{+}$ has EEE symmetry, and similarly $\Psi_{-}$
 has EEO symmetry.
When one compares with the numerical results of Martens et al \cite{9}, the estimates (4.22) are qualitatively correct;
the error of $E_{+}$ is about $+3.7\%$ while the error of $E_{-}$ is $+6.4\%$. One can further apply the factor
of $23/24$, obtained in the previous section, to end with the estimates 1.7471 and  1.556.
The result $E=1.7471$ for the (2,0) EEE state is
given in Table I as $E_{split}$. Other low-lying EEE state energies are also given.

    We end this brief excursion into the $x^2 y^2$ model with the conclusion that the quantum variational
method for excited states \cite{KN} is particularly well adapted for energy estimates in this system.
We have used variation-perturbation theory here to demonstrate this method. The same results can also be
obtained purely variationally, by using more elaborate trial states than (4.2); e.g., for an EEE state we would try (for
$n_x,n_y$   even)

$$
\psi_{n_x,n_y}^{(+)} (x,y) =N_{+}[ \psi_{n_x} (\omega_{x};x)\psi_{n_y} (\omega_{y}; y)+
 \psi_{n_y} (\omega_{y};x)\psi_{n_x} (\omega_{x}; y)],
$$
and similarly we would try $\psi_{n_x,n_y}^{(-)} (x,y)$ for the EEO states. We return to purely semiclassical variational 
methods in Sec. 4.3. For completeness, we note that classical variation-perturbation methods have also been  devised \cite{Raj}.

\subsection{The ``Linear Baryon"}

    Another system very amenable to classical and quantum
variational estimates is a system of three equal mass particles constrained to
move along a straight line \cite {L1}. There are constant attractive forces between
the particles, or a linear potential ( as for 1D gravity ), and the particles can move through each other. 
In suitable units the Hamiltonian can be
expressed as
$$
H=\frac{1}{2} [p_1^2 + p_2^2 + p_3^2] + \frac{1}{2}[|x_1- x_2| +|x_2 - x_3| +|x_3 - x_1|],  
\eqno(4.23)
$$
where $x_i$ are the positions and $p_i$ the momenta. If we keep the centre of mass at the origin we have the constraints 
$ x_1 + x_2 + x_3 = 0$ and $p_1+p_2+ p_3 = 0$. 
We can then take internal Jacobi coordinates $\rho,\lambda$ defined by
$x_1 - x_2 =\rho\sqrt {2}$ and $x_1 + x_2 - 2 x_3 = \lambda\sqrt{6}$, and the Hamiltonian becomes
$$
H= \frac{1}{2}[p_{\rho}^ 2 + p_{\lambda}^ 2] + \frac {1}{2\sqrt 2} [|\lambda\sqrt3 - \rho| + |\lambda\sqrt3 + \rho| + 2|\rho |].   
\eqno(4.24)
$$
   Taking now polar coordinates $R,\theta$  in the $\rho,\lambda$  plane, where $\lambda = R \cos\theta$ and $\rho = R \sin\theta$,  we obtain
$$
H=\frac {1}{2}[p_R^2 + p_{\theta}^2/R^2] + \frac{1}{\sqrt2} R(|\sin\theta|+|\sin(\theta + 2\pi/3)|+|\sin(\theta -2\pi/3)|).
\eqno (4.25)
$$

The equipotentials of this Hamiltonian form  regular hexagons.The
six sides correspond to the six different permutations of the three particles in
the baryon; for example, one side corresponds to $x_1 < x_2 < x_3$, etc.The
Hamiltonian (4.25) describes the motion of a particle inside this regular
hexagon. This system has been studied in the literature \cite {L2} under the
restriction to a single ``wedge" (say $0 < \theta < \pi/3$) under the name of `` wedge
billiards in a gravitational field". Recently also the relativistic generalization
of the Hamiltonian (4.23) has been discussed \cite{L3}.
    
The classical motion with the Hamiltonian (4.25) can be solved piecewise exactly.
The trajectory consists of segments of parabolas in each wedge of the
hexagon. One can visualize the system as a hexagonal ``pit" or ``funnel" in a
gravitational field , with a point particle sliding inside the funnel. Since the
system is not harmonic there are no normal modes, but an infinity of 
trajectories at each energy. Many of these can be simply described in terms of
two prototypes having high or low ``angular momentum" respectively. We
use quotation marks since angular momentum is not conserved in a
hexagonal potential. However the average angular momentum of specific
orbits is still meaningful. The low ``angular momentum" trajectory consists in
a motion along a groove of the hexagonal funnel corresponding to  $\theta = 0$, to
the origin and continuing along $\theta = \pi$, and back. The high ``angular
momentum" trajectory is composed of segments of parabolas joined together
to look like a circle. Each of these trajectories can sustain transverse
oscillations to give rise to quasiperiodic motions. Examples are shown in
references \cite{L1,L3}. Eventually for large transverse oscillations the two types
of trajectories blurr and and chaotic motion ensues. 

With only one constant of the motion (the energy) and two degrees of freedom, the system is nonintegrable, 
and possesses chaotic trajectories. These and the nonchaotic ones are piecewise parabolic, but it may be 
advantageous to represent them by simpler trial functions. For  illustration we
describe a low angular momentum trajectory  $\theta = p_{\theta} = 0$ when (4.25)
becomes 
$$
H= \frac{p_R^2}{2} + \frac {R \sqrt 3}{2}. 
\eqno  (4.26)
$$
Consider the trajectory with $E = 1$. The problem can be solved exactly to give a period (for the whole cycle) of 
$T_{exact}$ = 4.6188, while a variational estimate with the RMP using a harmonic oscillator
trial trajectory (with the same mean energy $\bar {E}$  = 1 as the exact trajectory) gives 
$T_{variational}$ = 4.6526, an overestimate of $0.73\%$.

\begin{figure}[h]
\begin{center}
\includegraphics[width=8cm]
{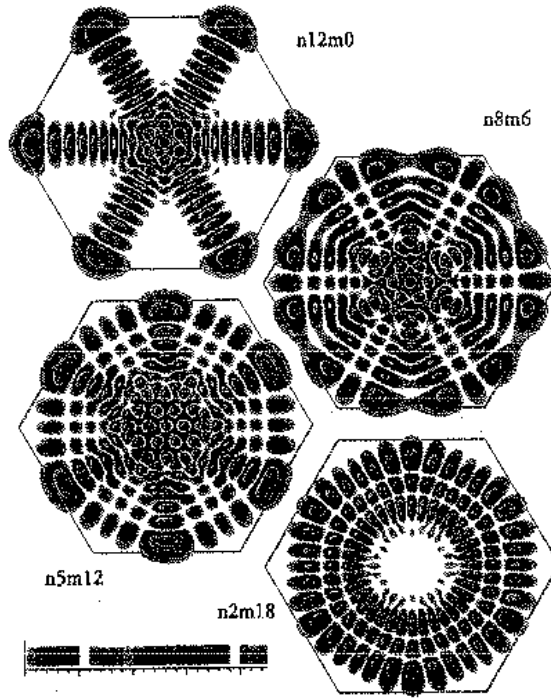}
\end{center}
\caption{The quantum probability distributions for four excited states of the hexagon pit. The notation $n12m0$ means the state $n=12, m=0$. The color scale of relative probabilities is given at the bottom. The graphs have been rescaled to make the hexagon, the classical boundaries of the orbits for the given $E$, the same size.(from ref. \cite{L1}) }
\label {fig:4}
\end{figure}

Quantum variational methods are quite useful here too since we cannot 
solve the quantum problem exactly, either for the ground state or excited
states. To describe the symmetry species of the  symmetry group 
C$_{6v}$ of the quantum Hamiltonian we can
use eigenfunctions of the 2D isotropic harmonic oscillator  \cite{L4}. In polar
coordinates these are products of an exponential $\exp(-\omega R^2/2)$ times a
function of $\theta$  ( $\sin(m\theta)$ or $\cos(m\theta)$) times $(\omega R^2)^{m/2}$ times a Laguerre
polynomial in $\omega R^2$ times a normalization factor. Although the azimutal
quantum number $m$ is not conserved with the Hamiltonian (4.25) it is
conserved modulo 6 due to the hexagonal symmetry of the problem. We can as a first
approximation minimize the diagonal elements of the Hamiltonian (4.26) in
this basis with respect to the frequency   and obtain first order estimates for
the energies of states which are labelled by a principal quantum number n
and m(modulo 6). For example  the energy of the ground state (n,m)=(0,0) is overestimated by
about $0.39\%$ in this basis while energy of the state (n,m)=(15,0) is overestimated by
about $3\%$. ``Exact" energies can be obtained by diagonalizing the
Hamiltonian in a large basis of harmonic oscillator states and minimizing
eigenvalues with respect to $\omega$. Another variational procedure uses products of
Airy functions along the groove and perpendicular to it, and takes the scale
factors of the two Airy functions as variational parameters. For the excited
state (15,0) the error is $0.014\%$  (the ``exact" energy is 16.34869 while the
Airy approximation gives 16.3511). More examples can be found in Table 1 of
reference \cite{L1} . The probability densities of a few states are shown in Fig. 4  
where one can see one low angular momentum state (12,0) and one high angular
momentum state (2,18) and two states which presumably correspond to
chaotic trajectories.

\subsection{He Atom Inside C$_{70}$ Cage}

Endohedral fullerene complexes such as He @ C$_{70}$ have attracted much study in recent years, both 
experimental and theoretical (for reviews see \cite {42a, 42b, 42c, 42d, 42e}). Here X @ C$_n$ denotes 
species X (atom, ion or molecule) encapsulated in a C$_n$ cage \cite {42f}.
 Early theoretical work focussed on the prediction of infrared and Raman \cite {42i, 52a} and neutron 
scattering \cite {42e} spectra due to the motion of X = He, Li$^{+}$, Na$^{+}$, K$^{+}$ and CO inside 
the near-spherical C$_{60}$ cage. The predictions of \cite {42i}
 were subsequently confirmed experimentally \cite{42j, 53a} for
the infrared bands of Li$^{+}$@C$_{60}$. Confirmation of the predicted Raman bands \cite{42i} of Li$^{+}$@C$_{60}$ is tentative \cite {53a}.
 Later the theoretical studies were extended to study the dynamics and spectra of atoms, molecules
 and ions trapped in nonspherical cages, e.g. He @ C$_{70}$ \cite{42k, 42l, 42m}, Ne @ C$_{70}$ \cite{42n}, He @ C$_{250}$ (nanotube) \cite {42o},
 Na$^+$ @ C$_{120}$ (nanotube) \cite{Cio}, and even (C$_{60}$)$_n$ @ C$_{\infty}$(nanotube), so-called nanopeapods \cite {59a}.
 Figs 5 and 6 show C$_{60}$,  C$_{70}$ and carbon nanotube cages.
\begin{figure}[h]
\begin{center}
\includegraphics[width=7cm]
{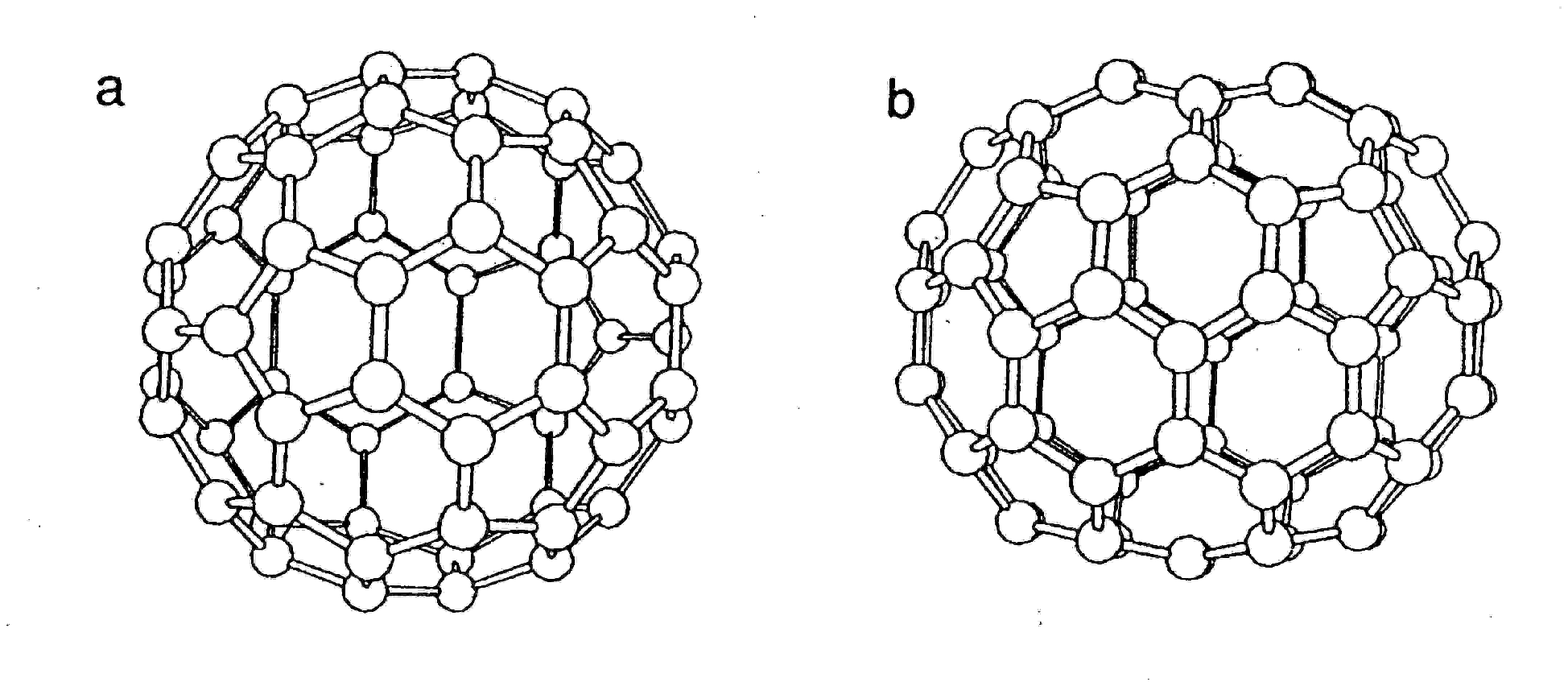}
\end{center}
\caption{(a) The soccer ball-shaped molecule C$_{60}$ (icosahedral symmetry group I$_h$); (b) The rugby ball-shaped molecule C$_{70}$ 
(symmetry group D$_{5d}$). (from ref. \cite{42b})}
\end{figure}
\begin{figure}[h]
\begin{center}
\includegraphics[width=7cm]
{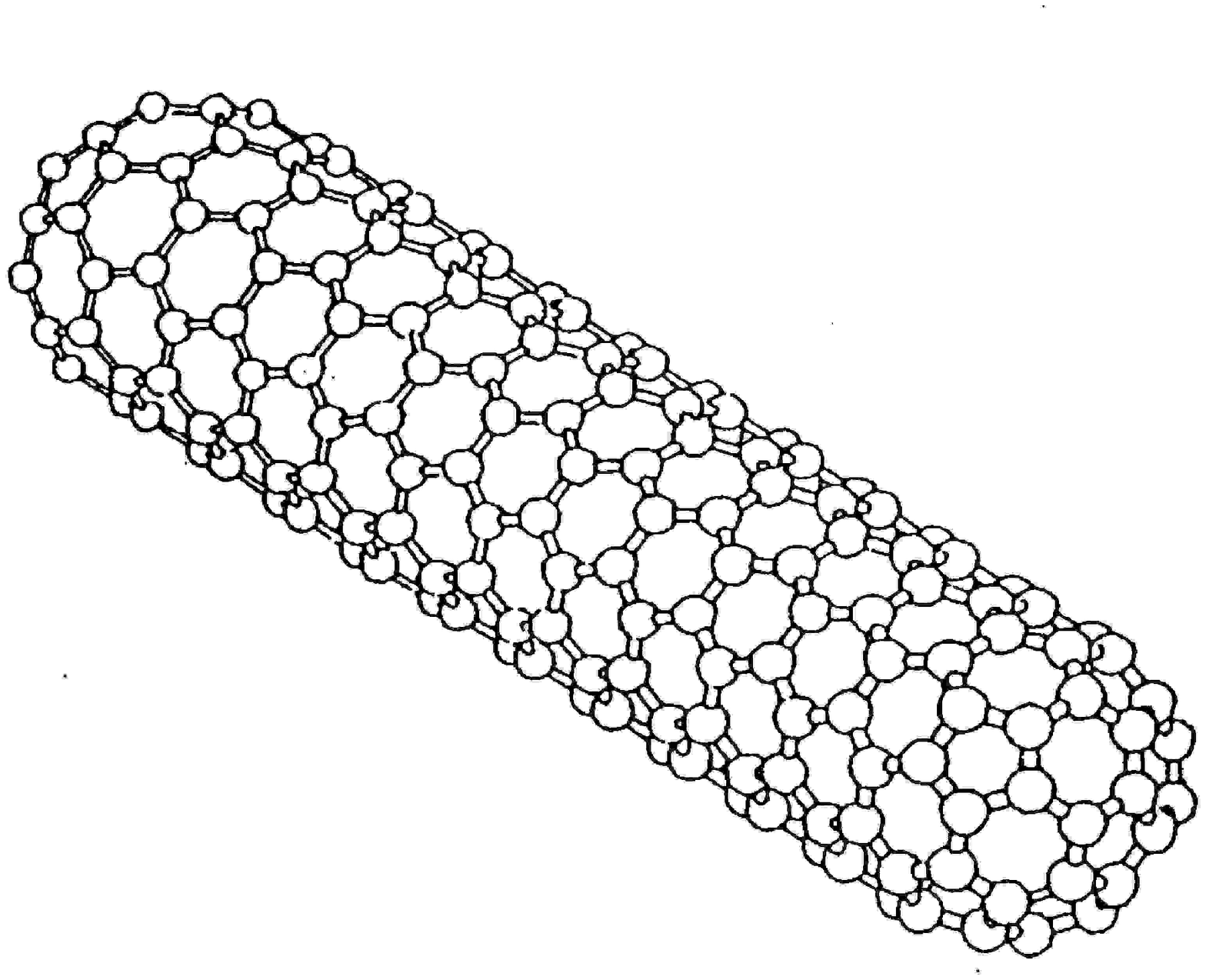}
\end{center}
\caption{Carbon nanotube of diameter 0.7 nm. This is a quite narrow nanotube with the hemispherical endcaps derived from C$_{60}$. 
The molecule depicted is C$_{250}$; actual nanotubes are much longer \cite{Yak}, up to C$_{1,000,000}$ and beyond. (from ref \cite {42o})}
\end{figure}
\\

The C$_{60}$ cage, assumed to be rigid,  presents a very nearly spherical potential \cite {59b}
 for an entrapped atom or ion X and the dynamics has been studied by various classical \cite {42p}, 
semiclassical \cite {42d} and quantum \cite{42d} methods. The potential can be calculated by a variety of semiemperical and {\it ab initio} 
methods \cite {42c, 42d, 42p, 42pa}.  The dynamics is usually quite regular \cite {59b} for X@C$_{60}$. We discuss here instead the case 
He @ C$_{70}$, where the potential is highly non-spherical and gives rise to nonintegrable (chaotic) dynamics of the He atom \cite {42k, 42l, 42m}. 
We estimate the energy levels semiclassically \cite{42l, 42m} using the UMP as described earlier, and compare with the ``exact" results 
obtained by brute-force diagonalization of the Hamiltonian \cite{42k}. Classical simulations of the motion have also been carried 
out \cite{42k, 42l, 42m}, and provide insight as to the degree of chaos in the motion of the He atom. The chaos is surprisingly small, 
given that the potential is strongly anharmonic. 
The small degree of chaos implies an approximate third constant of the motion  \cite {42q}, in addition to the energy and $z$-component of 
angular momentum (assuming the potential is axial).

 The potential for He in a rigid C$_{70}$ cage has been estimated semiemperically \cite{42k}, and, to a good approximation, 
is axially symmetric around long $z$-axis in Fig 5b . With origin at the centre of the cage and $x,y$ axes perpendicular to 
the symmetry $z$-axis, the potential has been fit to the form

$$
V(x,y,z)= \frac{1}{2} k_1 (x^2 + y^2) + \frac{1}{2} k_3 z^2 + C (x^2 + y^2)z^2.
\eqno (4.27)
$$   
The first two terms describe a 3D anisotropic harmonic oscillator, with frequencies 
$~\omega_x^{(0)} =\omega_y^{(0)}= (k_1/m)^{1/2}= 84.166~  cm^{-1}$ and 
$\omega_z^{(0)}=(k_3/m)^{1/2}= 21.934 ~ cm^{-1}$, where $m$ is  the He atom mass. We use the standard spectroscopic 
unit ($cm^{-1}$) for frequency and energy (see \cite{42l}). The last term $C (x^2 + y^2)z^2$ in (4.27) is a
quartic perturbation, with  $C=1.738 ~ cm^{-1}$ when $x, y ,z$ are in units of $(\hbar/m \omega^{(0)}_x)^{1/2}$. 
There are also smaller terms in (4.27) varying as $(x^2 + y^2)^2$, $z^4$, etc, which we shall ignore.

Because of the axial symmetry we change to cylindrical polar coordinates ($\rho, \phi, z)$ and write the Hamiltonian, with potential (4.27), as 
$$
H =  \frac{1}{2} m({\dot \rho}^2 + {\dot z}^2) + \frac{l_z^2}{2m\rho^2}+\frac{1}{2} m{\omega_x^{(0)}}^2 \rho^2 +\frac{1}{2} m{\omega_z^{(0)}}^2 z^2 +  C \rho^2 z^2,
\eqno(4.28)
$$
where $\rho = (x^2+y^2)^{1/2}$, and $l_z$ is the z-component of the angular momentum, which is a constant of the motion.
 The cage has been assumed to be rigid and fixed in space.
 To apply the UMP (2.10), we proceed exactly as for the simpler spherical pendulum example of sec.3.2. 
When viewed down the $z$ axis, the trajectory is roughly a precessing ellipse, and the $z$-motion is oscillatory. 
We therefore take as a trial trajectory
$$ 
x^{'}= A_x \cos \omega_x t,\;\;\  y^{'}= A_y \sin \omega_x t,\;\;\ z^{'}= z = A_z \cos \omega_z t,
\eqno (4.29)
$$
where the $x^{'}, y^{'}$ axes precess with some (unknown) frequency $\Omega$ with respect to $x, y$, and where $\omega_x$ may differ from $\omega_x^{(0)}$, etc. Calculating $\bar E$ and $W$ with the trial trajectory and applying the UMP gives to first order in $C$ (see \cite {42l} for the details) $\Omega = 0$, the shifted frequencies
$$
\omega_x = \omega_x^{(0)} + \frac {C W_z}{2\pi m^2 \omega_x^{(0)}\omega_z^{(0)}}, \;\;\;\
\omega_z=\omega_z^{(0)} + \frac {C (W_x + W_y)}{2\pi m^2 \omega_x^{(0)}\omega_z^{(0)}},
\eqno (4.30)
$$
and also, to first-order in $C$, $\bar E$ as a function of the partial actions $I_i = W_i/2\pi$
$$
\bar E(I_x,I_y,I_z)= (I_x + I_y)\omega_x^{(0)} + I_z \omega_z^{(0)} +  \frac {C (I_x + I_y)I_z}{ m^2 \omega_x^{(0)}\omega_z^{(0)}}.
\eqno(4.31)
$$

The terms of second-order in $C$ in (4.30) and (4.31) are also derived in \cite{42l}, and it is also found that $\Omega = 0$ to second order. The simulations show that $\Omega$ is small but nonzero in general, and to obtain $\Omega\not = 0$ from the UMP requires an improved trial trajectory over (4.29), e.g. one involving higher harmonics of $\omega_x$ and $\omega_y$ \cite{42m}.
We shall ignore this small effect here. 

To estimate the energy levels semiclassically from (4.31) is very simple: according to the EBK quantization rule (2.11), we replace $I_x$ by $(n_x + \frac{1}{2})\hbar$, etc. This gives 
$$
 E_{n_x,n_y,n_z}= (n_x + n_y+1)\hbar \omega_x^{(0)} + (n_z+ \frac {1}{2}) \hbar \omega_z^{(0)} +  
\frac {C (n_x + n_y+1)(n_z + \frac{1}{2})\hbar^2}{ m^2 \omega_x^{(0)}\omega_z^{(0)}}+ O(C^2)
\eqno(4.32)
$$      
where the $O(C^2)$ and $O(C^3)$ terms are written down in ref. \cite {42l}. We compare the semiclassical energy levels (4.32)  
(including the   $O(C^2)$ and $O(C^3)$ terms) with the fully quantum results; the lowest 613 levels, up to the energy of 
$1000~cm^{-1}$ are obtained by diagonalizing the Hamiltonian in a harmonic oscillator basis in \cite {42l}. The harmonic oscillator 
chosen has a potential given by the harmonic part of (4.28). The semiclassical and fully quantum
results agree to better than $1\%$.
We illustrate the good agreement in Fig 7, which shows the levels up to $500~ cm^{-1}$, organized according to $m$, 
the quantum number corresponding to $l_z$.

\begin{figure}[h]
\begin{center}
\includegraphics[width=10cm]
{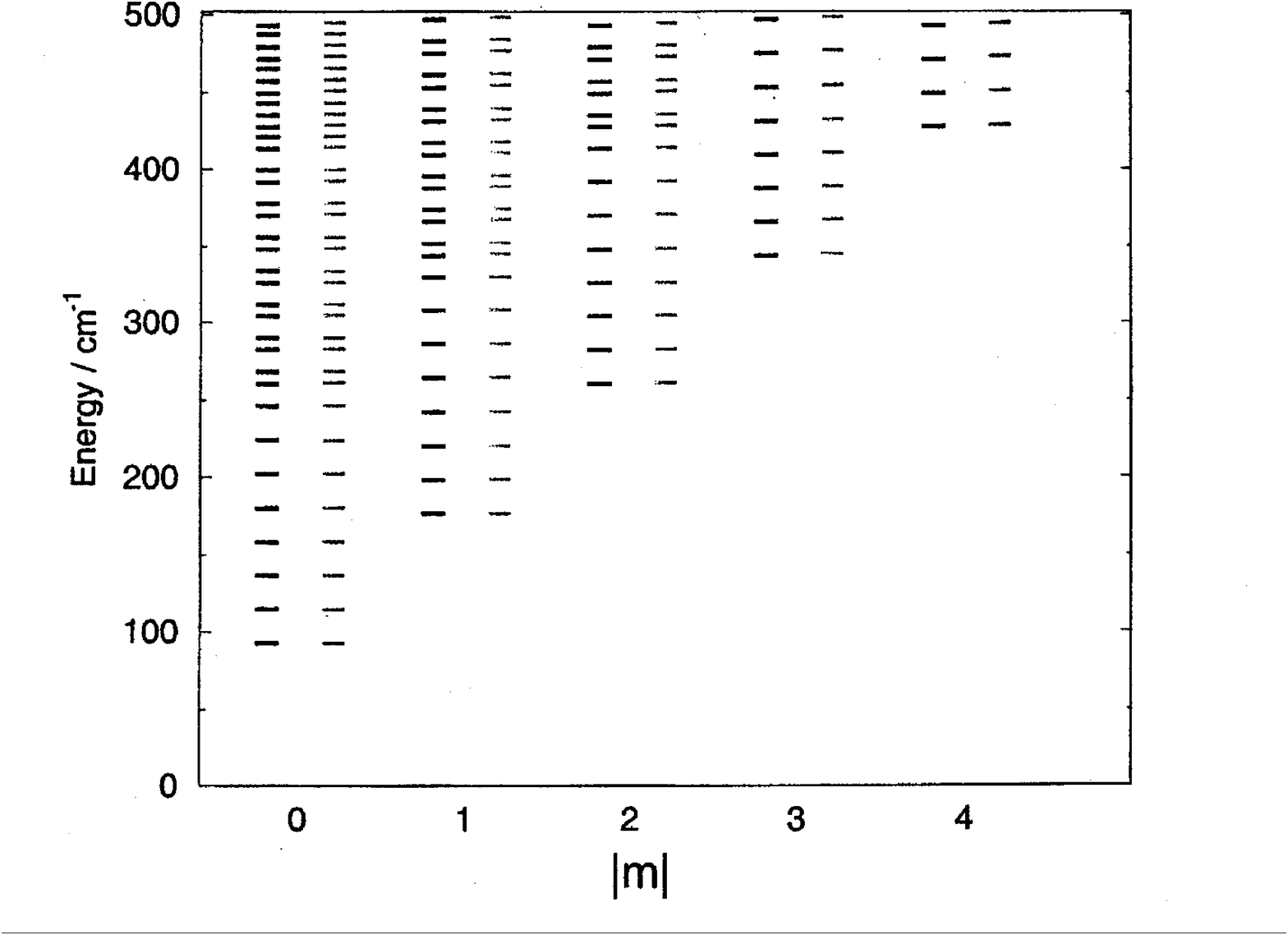}
\end{center}
\caption{Comparison of energy levels calculated via quantum mechanical methods (black) and a semiclassical 
variational technique (grey).(from ref.\cite{42l})}
\end{figure}
For an isotropic 2D harmonic oscillator, the energy depends only on $N_{xy}= n_x + n_y$. For $N_{xy} = 2$, for example,
 we can have $(n_x,n_y) = (2,0), (1,1), (0,2)$, which are degenerate. The $m$-values allowed for $N_{xy}=2$ are $m=-2, 0, +2$. 
For general $N_{xy}$, we have $|m|= N_{xy}, N_{xy}-2, N_{xy}-4, ...,1$ or $0$.
For our oscillator (4.28) the different $|m|$ states should in reality have different energies; this is found in the fully quantum calculation, but not in the semiclassical one where the energy depends only on  $N_{xy}= n_x + n_y$. Table 2 illustrates more clearly the slight $|m|$ splittings found quantum mechanically, but not semiclassically.

To obtain the $|m|$ splittings semiclassically requires an improved trial trajectory over (4.29), one that gives a nonvanishing precession frequency $\Omega$. This is discussed in ref \cite {42m}. The simple improvement in the trial trajectory mentioned earlier (including some higher harmonics) does indeed give a nonvanishing $\Omega$ and hence $|m|$ splitting. However the trial trajectories employed thus far are unable to give \underline{accurate} splittings, and hence require further improvement. 

\vspace{5mm}
\begin{center}
{\bf Table 2}
\vspace{5mm}

\begin{tabular}{|c|c|c|c|c|c|c|}
\hline

$n_z $ & $N_{xy}=n_x+n_y$& $ m $ & $E_{quant}(cm^{-1})$ & $ E_{semiclass}(cm^{-1})$\\
\hline
 0  &  2  &   2   &  262.818   &   263.466 \\

 0  &  2  &   0   &  262.821   &   263.466 \\

 1  &  2  &   2   &  284.696   &   285.403 \\

 1  &  2  &   0   &  284.699   &   285.403 \\

 2  &  2  &   2   &  306.574   &   307.340 \\

 2  &  2  &   0   &  306.577   &   307.340 \\

\hline
\end{tabular}

\end{center}

{ Comparison of selected energy levels calculated via quantum and semiclassical methods demonstrating the failure of the simple variational calculation to produce level splitting
observed in the quantum results.(from ref.\cite{42l})}

\section{Nonconservative Systems}

 In this section we consider nonconservative mechanical systems. For
example consider a system of particles moving in the time-dependent
external field, or a system whose parameters depend explicitly on time.
Other examples include time-dependent constraints, e.g. a bead  on a 
rotating hoop whose angular velocity is a prescribed function of
time, and variable mass systems (for which $L\neq K-V$ in general
\cite{Leu}).
Such systems
are not closed and the energy is not conserved. The evolution of
a nonconservative system is
completely determined by a time-dependent Lagrangian  $L(q,\dot{q},t)$ or by
a time-dependent Hamiltonian $H(q,p,t)$.
(We exclude dissipative  systems from our discussion. We do note
that Lagrangians and Hamiltonians have been found for such systems
in some cases \cite{Bat}, so that the action principles discussed
 here would apply in these cases. Alternatively \cite{1b}, dissipative terms can be added
to the Lagrange equations of motion or to the Hamilton principle.) $H$ and $L$ are related to each other by
the Legendre transformation (see eq. (2.5))

$$
  H(q,p,t) = \sum_n p_n \dot{q}_n  - L(q,\dot{q},t).
  \eqno(5.1)
$$
Locally these lead to the usual Euler-Lagrange or Hamilton equations of motion that govern the
dynamics of the system.

\subsection{Action Principles}

We consider the global description of nonconservative systems.
It is well known that the Hamilton Principle
$$
(\delta{S})_T = 0
\eqno(5.2)
$$
is applicable for nonconservative systems as well as for conservative ones. Here subscript $T$
reminds us that variations near the actual path are constrained by fixing
the travel time $T$ from the initial point $A\equiv {(q_A, 0)}$ to the final point $B\equiv {(q_B, T)}$. Again we have chosen
 $t_A=0 $ and $t_B=T$, and, as before,
 it is understood here and below that the end-positions $q_A$ and $q_B$ are also fixed.

As before, we can relax the constraint of fixed $T$  and rewrite (5.2)\cite {5, Buc} as an unconstrained  Hamilton Principle (UHP)
$$
\delta{S} + E(T)\delta{T} = 0,
\eqno(5.3)
$$

\vspace{2mm} \noindent
where the Lagrange multiplier $E(T)$ is the true value of the energy at the time of arrival at
the final point $ B$. (Elsewhere in this paper $E(T)$ is denoted variously as $E(t_B)$, $E_B$ or $H_B$; see notes \cite{note2} and \cite{note3}).
 We assume that all trial trajectories start at the same time $t= 0$ and also have the same end positions $q_A$ and $q_B$.

 Using eq.(2.7), which is derived from (5.1) as before, we transform the UHP (5.3) to the form

$$
T\delta{\bar E} - \delta{W} - [ E(T)- \bar E(T)]\delta{T} = 0.
\eqno(5.4)
$$

\vspace{2mm} \noindent This is the unconstrained  Maupertuis Principle (UMP) for nonconservative systems,
which seems not to have been given in the
literature explicitly (it is implicit in an Appendix in \cite{5} - see also \cite{note3}). It relates
three variations - the variation of mean energy $ \bar E$,
the  variation of the action $W$
and the variation of the travel time  $T$. The Lagrange multipliers
are the true travel time
$T$ and the difference of energy and mean energy of the true trajectory at the final point,
$ [ E(T)- \bar E(T)] $. The latter can be rewritten as

$$
E(T)- \bar E(T) =\frac {1}{T}\int\limits_0^T  \ {dt} [ {t} \frac{\partial H}{\partial t}]=
<t\frac{\partial H}{\partial t}>,
\eqno(5.5)
$$
where the bar and brackets $<...>$ both denote a time average over the interval (0,T).
The generalized UMP (5.4)
can also be derived by traditional variational procedures (\cite{5}, eq.(A.6)).

One can derive constrained principles from (5.4) as before, e.g. $(\delta{\bar E})_{W,T} = 0$,
 but these would be complicated to use in practice because of the severe constraints on the
 trial trajectories. The version $(\delta{W})_{\bar E, T}=0$ is closest to the original
 Maupertuis Principle. The constrained principle
 $$
 \left(\delta[W+T<{t} \frac{\partial H}{\partial t}>]\right)_{E(0),E(T)}=0
 $$
  is mentioned by
 Whittaker \cite{Whit}, and is easily derived from (5.4) and (5.5). But since (5.4) is also 
valid without fixing $E(0)$ and $E(T)$ , (5.4) is more convenient.

Formally the unconstrained  Maupertuis principle (5.4) is equivalent to the Hamilton principle,
but it also appears too complicated to be useful in practice.
This is not quite true and we demonstrate how the UMP can be used for  adiabatic perturbations.

\subsection{Adiabatic Invariants and the Hannay Angle}

For simplicity we start with a one-dimensional periodic system that depends on a parameter ${\lambda}$.
Let ${\lambda}$ vary with time adiabatically, i.e. the change 
${\Delta\lambda}\ll {\lambda}$ during a period of oscillation $T$. It is well known that in this case
the energy $E$
as well as the period of oscillation $T$ depend slightly on time, but the action $I$ over a cycle, where

$$
I = \frac {1}{2\pi}W(cycle)=\frac {1}{2\pi}\oint\ p{dq},
\eqno(5.6)
$$

\vspace{2mm} \noindent
remains constant to an excellent approximation (adiabatic invariant) when the parameter ${\lambda}$ varies.

 Consider what kind of information  can be extracted from the
 UMP (5.4).  Both the energy $E$ and the period $T$ of an actual motion are
functions
of the action $I$ (recall the action-angle variables): $E=E(I)$, $T=T(I)$.
Consider two actual trajectories with actions $I$ and $I+dI$ as two
 trial trajectories for the UMP (5.4). Then the variational equation can be rewritten
 as a differential equation

$$
\frac{d\bar{E}}{dI} - \frac{2\pi}{T}= <t\frac{\partial E}{\partial t}>\frac{1}
{T}\frac{dT}{dI}.
\eqno(5.7)
$$

\vspace{2mm} \noindent
It is more convenient to rewrite this in terms of the frequency
$\omega(I)={2\pi}/{T(I)}$,
$$
\omega(I) = \frac{d\bar{E}}{dI} + <t\frac{\partial E}{\partial t}>\frac{1}
{\omega(I)}\frac{d\omega(I)}{dI}.
\eqno(5.8)
$$

\vspace{2mm} \noindent
This equation generalizes the standard relation \cite {1b} for integrable conservative systems
$$
\omega(I) = \frac{dE}{dI}
$$
to integrable nonconservative systems.

Up to this point (5.7) and (5.8) are  general and valid for any nonconservative system for which a good action
can be defined.
Consider now the adiabatic regime where  the variation of the parameter $\lambda(t)$ over the short
 period $T$,  $\Delta\lambda= \lambda(T) -  \lambda(0)$,
is much less than the parameter itself:   $\Delta\lambda \ll \lambda(0)$. In this case
$$
<t\frac{\partial E(I,\lambda(t))}{\partial t}>=
<t\frac{\partial E(I,\lambda)}{\partial \lambda}\dot\lambda> =
\dot\lambda<t\frac{\partial E(I,\lambda)}{\partial \lambda}>,
\eqno(5.9)
$$

\vspace{2mm} \noindent
where the averaging $<t{\partial E(I,\lambda)}/{\partial \lambda}>$
is to be performed at a fixed value of $\lambda\simeq \lambda(0)$ since $\lambda$ varies little over a period $T$.
 The action $I$ is
conserved for any given parameter $\lambda$, and hence the derivative
${\partial E(I,\lambda)}/{\partial\lambda}$ is also a constant of the
motion.
As a result in this approximation we get

$$
<t\frac{\partial E(I,\lambda)}{\partial \lambda}>=
 \frac{T}{2}\frac{\partial E(I,\lambda)}{\partial \lambda}= \frac{\pi}{\omega}\frac{\partial E(I,\lambda)}{\partial \lambda}.
$$

\vspace{2mm} \noindent
In the same approximation we have for $\bar E\equiv <E>$ in (5.8)
$$
< E(I,\lambda(t))>= E(I,\lambda(0)) +\frac{\pi}{\omega} \dot\lambda \frac{\partial E}{\partial \lambda}.
$$

Thus to the first order approximation in $\dot\lambda$ we can rewrite  (5.8) as

$$
\omega(I) = \frac{d{E(I,\lambda(0))}}{dI} +
[\pi \dot\lambda][\frac {d}{dI} (\frac{1}{\omega^{(0)}}\frac{\partial E}{\partial \lambda})-
 \frac{\partial E}{\partial \lambda}\frac{d}{dI} (\frac{1}{\omega^{(0)}})],
$$
or finally as

$$
\omega(I)= \omega^{(0)}+ \frac{\Delta\lambda}{2}\frac{\partial \omega^{(0)}}{\partial \lambda}.
\eqno(5.10)
$$
Here $\omega^{(0)}(I,\lambda)=dE(I,\lambda(0))/dI$
is the zero-order term, and $\Delta\lambda=\lambda(T)-\lambda(0)\simeq \dot\lambda (2\pi/\omega^{(0)})$ is the variation of the parameter $\lambda$
during the given period.
 We get corrections (of the first-order in the nonadiabaticity parameter $ \dot\lambda$) to the
frequency of oscillation, for adiabatic systems. The result  seems very simple. It appears that $\omega(I)$ is a
continuous  function of
time $t$ and we calculate some average frequency over the period. But actually  $\omega(I)$ is not always a continuous
function of time;
it is  a \underline{global} parameter of  a (slightly) aperiodic system. Thus  (5.10)  is nontrivial
 in general.

 To illustrate these results it is instructive to consider a simple exactly solvable example. We consider the 1D
 motion of a perfectly
 elastic  ball of mass $m$ bouncing between two planes. We suppose that one plane is fixed and the other is slowly
 moving with constant velocity $u$. We suppose also that the planes are heavy enough so that they are unaffected
 by collisions with the ball.
  The distance between the two planes  $ L(t) = L_0 + ut$  is the  adiabatic parameter.

  Let us define a \underline {cycle} as the motion of the ball from the fixed plane  to the moving plane, collision, and
  the motion of the ball in the  backward direction up to the collision with the fixed plane.
The n-$th$ cycle can be described in the following way. At time $t=0$ (beginning of the cycle) the ball
has  velocity  $v_n$, and the distance between planes is $L_n$.  At time $T_n(\rightarrow)=L_n/(v_n-u)$ the ball
collides with the moving plane and after the collision
  it moves to the fixed plane with velocity $(v_{n}-2u)=v_{n+1}$ (the collision with the fixed plane
  is elastic and the backward velocity is equal to
  the forward velocity after collision, i.e. $v_{n+1}$).
 The backward motion takes time $T_n(\leftarrow)=(L_n v_n)/(v_n-u)v_{n+1}$.
 The period of the whole  cycle is $T_n= T_n(\rightarrow)+T_n(\leftarrow)= (2L_n)/v_{n+1}$.
At time $t=T_n$ (the initial time for (n+1)-th cycle) the distance between the two planes is equal to $L_{n+1}= L_n v_n/v_{n+1}$.
This means that  $L_nv_n=L_{n+1}v_{n+1}$  is an exact (adiabatic) invariant.

It easy to see that the action $I$ is proportional to this invariant,
$$
I_n =\frac {1}{2\pi}\oint\ p{dq}=\frac{m}{\pi}L_nv_n,
$$
i.e. the action $I$ is an exact (adiabatic) invariant for our special initial condition \cite{note5}.

In terms of the action $I$ and the adiabatic parameter $L$ the complete set of  relevant quantities is as follows:

$$
\bar E_n=\frac{1}{2m}(\frac{\pi I}{L_n})^2-u(\frac{\pi I}{L_n}),
$$
$$
E(T_n)=\frac{1}{2m}[(\frac{\pi I}{L_n}) - 2m u]^2,
$$
$$
\omega_n(I)= \frac{2\bar E_n}{I}=\frac{1}{2m}(\frac{\pi^2 I }{L_n^2})-u(\frac{\pi }{L_n}),
$$

\vspace{2mm} \noindent
i.e. $\omega_n(I)$ is linear function of the nonadiabaticity parameter $\dot L=u$ and the
approximate relation (5.10) must be exact here. Indeed, we have
$\omega^{(0)}=\pi^2 I/2mL_n^2$, $\Delta L=\pi u/\omega^{(0)}$ and
$$
\omega^{(0)}+\frac{\Delta L}{2}\frac{\partial \omega^{(0)}}{\partial L} =\frac{1}{2m}(\frac{\pi^2 I }{L_n^2})-u(\frac{\pi }{L_n})=\omega(I).
$$
Thus the approximate relation (5.10) is valid here.  It is also not difficult to check after a little algebra that in the
exact relation (5.8) all higher powers of $u$ cancel each other and  the  linear approximation is exact.

If $u\equiv \dot {L}$ is not extremely small, $I$ ceases to be an adiabatic invariant and the
system becomes nonintegrable. If $L(t)$ is varied nonadiabatically and periodically around $L_0$, 
the system  (termed the Fermi model \cite{64a})  can show steady energy growth and chaotic motion.

The generalization of (5.10) to multidimensional integrable systems with
actions $I_1$,...,$I_f$ is evident, as is the generalization to multiparameter systems, with
parameters $\lambda_1, \lambda_2, ... $.

The $O(\dot\lambda)$ corrections in $\omega(I)$ have been  derived
by other methods \cite{Han}, but in different forms from (5.10). If we
assume $\lambda(t)$ is periodic, with a long period $T_\lambda$,
the $O(\dot\lambda)$ terms in $\omega(I)$, when integrated over
$(0, T_\lambda)$, define the \underline {Hannay angle} $\Delta\theta_H$, the
classical analogue of the quantum Berry phase \cite{Han}. The Hannay angle is
the extra (geometric) phase undergone by the system in one cycle
of $\lambda(t)$, in addition to the more obvious dynamical phase
$\theta_D$ obtained by integrating the instantaneous frequency
$\omega(\lambda(t)) \equiv dE(I, \lambda(t))/dI$. In general one requires
more than one $\lambda$-type parameter to
obtain a nonvanishing result for $\Delta\theta_H$   , but Hannay and Berry and coworkers
give a few examples \cite{Han, 36a, 36b} where one parameter suffices.
We discuss two examples with nonvanishing $\Delta\theta_H$.

\subsubsection{Bead Sliding on Slowly Rotating Horizontal Ring}

Consider a bead of mass $m$ sliding without friction on a closed planar
wire loop which is slowly rotating with angular velocity $\Omega$ in the
horizontal plane. For a loop of arbitrary shape, Hannay and Berry
\cite {Han} show that the Hannay angle $\Delta\theta_H$ defined above is
given by $\Delta\theta_H = -8\pi^2 A/L^2$, where $A$ is the area
and $L$ the perimeter of the loop. For a circular loop rotating
about an axis perpendicular to the plane of the loop and through
the center, $\Delta\theta_H$ is simply $-2\pi$, as is clear
intuitively: the ``starting line'' for the particle moves ahead by
$2\pi$ while the particle rapidly slides around the loop.

\begin{figure}[h]
\begin{center}
\includegraphics[width=6cm]
{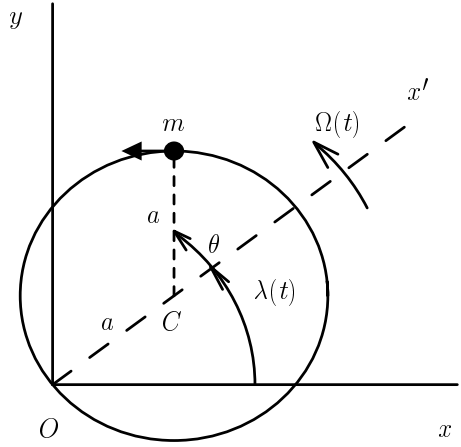}
\end{center}
\caption{Circular wire loop rotating slowly in the horizontal $x, y$ plane with angular velocity 
$\Omega=\dot\lambda$  about axis through $O$ perpendicular to the $x,y$ plane.
 At $t=0$, the particle starts with $\theta=0$ and the rotating $x'$ axis ( a diameter) coincides with the fixed axis $x$.}
\label {fig:8}
\end{figure}

We consider a circular loop of radius $a $ (Fig.8), but with the vertical
axis of rotation through a point $O$ on the wire instead of the center $C$ 
of the loop, to make
the answer less obvious. The adiabatic parameter here is the prescribed
orientation of the loop $\lambda(t)$, where $\dot\lambda(t) \equiv
\Omega(t)$  is small. We do not at this stage assume $\Omega(t)$ is
 constant.

Using the horizontal $x, y$ lab-frame (Fig.8), we can easily write the
Lagrangian, which is just the lab kinetic energy of the particle.
Using $\theta$ in Fig.8 as the generalized coordinate we find

$$
L = \frac{1}{2} ma^2 \left[\Omega^2 +(\Omega +\dot\theta)^2
+2\Omega(\Omega +\dot\theta)\cos\theta \right] \;\; .
\eqno(5.11)
$$
From (5.11) we obtain the equation of motion
$$
\ddot\theta +
\dot\Omega(1+ \cos\theta) +\Omega^2 \sin\theta = 0 \;\; .
\eqno(5.12)
$$
If the loop rotates uniformly ($\dot\Omega =0$), (5.12) is just the
pendulum equation, as one expects intuitively.

Because we are using a rotating coordinate system ($\theta$ is
measured from $x^\prime$) $L$ is not equal to $K-V\equiv  K =
\frac{1}{2} ma^2 \dot\theta^2$  in this system, so that the
Hamiltonian $H$ is not equal to the energy ($=\frac{1}{2} ma^2
\dot\theta^2$) in this system either. To find $H$, we first
calculate the canonical momentum $p_\theta$, where
$$
p_\theta = \frac{\partial L}{\partial \dot\theta} = ma^2(\Omega
+\dot\theta +\Omega\cos\theta) \;\; ,
\eqno(5.13)
$$
and then $H=p_{\theta}\dot\theta -L$ is
$$
H =\frac{1}{2} ma^2 \dot\theta^2
-ma^2 \Omega^2(1+\cos\theta) \;\; .
\eqno(5.14)
$$
In terms of the canonical variables $\theta$, $p_{\theta}$  the Hamiltonian  $H$
is
$$
H = \frac{p_\theta^2}{2ma^2} - p_\theta\Omega(1+\cos\theta) + O(\Omega^2) \;\; ,
\eqno(5.15)
$$
where we drop $O(\Omega^2)$ terms since we assume $\Omega\ll \dot\theta$.

We cannot apply (5.10) as it stands, since $H=E$ was used in the derivation. The generalization to
include cases where $H\not=E$ is
$$
\omega(I,\lambda) = \frac{d\bar{H}}{dI} - (\pi\dot\lambda \frac{\partial \bar{H}}{\partial \lambda})
\frac{d}{dI}(\frac{1}{\omega^{(0)}(I)} ),
\eqno(5.16)
$$
where $\bar H$ is the average of $H$ over the fast motion
(here $\theta$). Alternatively, we can use (5.10), with now
$\omega^{(0)} = dH(I, \lambda(0))/dI$.
From (5.15) we have
$$
\bar H \doteq \frac{p_{\theta}^2}{2ma^2} - p_{\theta}\Omega,
\eqno(5.17)
$$
since $\theta$ varies rapidly and $p_{\theta}$ is nearly constant.

We now assume $\Omega$ is constant. For $\Omega = 0$, $\bar H$ is
a function of the good action $I=p_{\theta}$ only (no $\theta$
dependence), and when $\Omega \neq 0$ we see $p_{\theta}$ remains
a good action. Note that $\bar H$ is independent of angle
$\lambda$, so that the second term in (5.16) will not contribute
to $\omega(p_{\theta}) = d\bar H/dp_{\theta}$, which is
$$
\omega(p_{\theta}) = \frac{p_{\theta}}{ma^2} - \Omega \;\; .
\eqno(5.18)
$$
The total phase undergone after a long return time  $T_\Omega =
2\pi/\Omega$ is thus
$$
\theta =(\frac{p_{\theta}}{ma^2}) T_\Omega -2\pi \;\; .
\eqno(5.19)
$$
The first term is the dynamical phase, the second is the Hannay
(geometric) phase.

\subsubsection{The Foucault Pendulum}

For a Foucault pendulum the Hannay angle is the angle by which the
plane of the pendulum vibration shifts in one day, i.e. $\Delta\phi_H =
-2\pi\cos\theta$, where $\theta$ is the co-latitude of the point
on earth where the pendulum is located. This is a clockwise
rotation in the northern hemisphere. This result for
$\Delta\phi_H$ has been derived by Berry \cite{36b} and others  \cite{36c}
by various arguments. There are also some pre-Hannay purely
geometric arguments for the Foucault pendulum precession  \cite{36c'}.
Here we show the result  follows simply from (5.16).

We consider a pendulum of length $L$ and bob mass $m$ oscillating
at a point on the earth's surface with co-latitude $\theta$. We
use axes x,y,z attached to the earth, with x,y in the horizontal plane and z
vertical. When at rest in the x,y,z system, the pendulum hangs along
the z-axis, and when in motion the bob oscillates very nearly in the
x,y plane. Neglecting the z motion and assuming $\Omega \equiv
\Omega_E \cos\theta$ is small, where $\Omega_E$ is the earth's
angular velocity, we easily find \cite{36d} the approximate horizontal equations of motion
$$
\ddot x + \omega_0^2 x -2\Omega \dot y = 0 \; , \;\; \ddot y +
\omega_0^2 y +2\Omega \dot x = 0 \;\; ,
\eqno(5.20)
$$
where $\omega_0 = (g/L)^{1/2}$. We have kept the $O(\Omega)$ Coriolis
terms, but dropped the $O(\Omega^2)$ centrifugal terms, since $\Omega$
is small compared to the pendulum frequency $\omega_0$. Here $\Omega$ is the local vertical
component of the earth's angular velocity. We are interested in particular
in the orbit precession rate predicted by (5.20).

It is easily checked that (5.20) follow from the Lagrangian
$$
L = \frac{1}{2}m(\dot x^2 +\dot y^2) -\frac{1}{2}m\omega_0^2
(x^2 +y^2) +m\Omega(x\dot y - y\dot x) \;\; ,
\eqno(5.21)
$$
and of course (5.21) can be derived directly \cite{LL1}.

Changing to polar coordinates $\rho$, $\phi$ via $x
=\rho\cos\phi$, $y = \rho\sin\phi$ changes $L$ to the form $$ L =
\frac{1}{2}m\dot\rho^2 +\frac{1}{2}m\rho^2\dot\phi^2 -
\frac{1}{2}m\omega_0^2\rho^2 +m\Omega\rho^2\dot\phi \;\; .
\eqno(5.22) $$
Two constants of the motion are obvious from (5.22): since $L$
does not depend on $\phi$, $p_\phi$ is a constant, where
$$
p_\phi =\frac{\partial L}{\partial\dot\phi} = m\rho^2 \dot\phi
+ m\rho^2 \Omega \;\; ;
\eqno(5.23)
$$
and because $L$ does not depend on $t$ explicitly, the
Hamiltonian $H = p_\rho \dot\rho + p_\phi \dot\phi -L$ is a 
constant, where $p_\rho = {\partial L}/{\partial \dot\rho}$ and
$$
H = \frac{1}{2}m\dot\rho^2 +\frac{1}{2}m\rho^2\dot\phi^2 +
\frac{1}{2}m\omega_0^2 \rho^2 \;\; .
\eqno(5.24)
$$
Two constants of the motion with two degrees of freedom imply that 
the system is integrable and that two good actions exist.

To express $H$ in terms of the canonical variables $\rho$, $\phi$,
$p_\rho$, $p_\phi$ we use (5.23) and
$$
p_\rho = \frac{\partial L}{\partial{\dot\rho}} = m\dot\rho
\eqno(5.25)
$$
to get
$$
H = \frac{p_\rho^2}{2m} +\frac{p_\phi^2}{2m\rho^2} +
\frac{1}{2}m\omega_0^2\rho^2 -\Omega p_\phi,
\eqno(5.26)
$$

\vspace{2mm} \noindent
where again we drop an $O(\Omega^2)$ term.
 We wish to use (5.16) (not (5.10) here as $H\neq K+V$) to find the
frequencies $\omega_\rho \equiv \omega(I_\rho, \lambda)$ and
$\omega_\phi =\omega(I_\phi, \lambda)$, where $\lambda$ is the
orientational angle of the earth multiplied by  $\cos\theta$. Note that for $\Omega =
\dot\lambda = 0$, (5.26) is simply the Hamiltonian for an
isotropic 2D harmonic oscillator. Calculation of the actions for
the 2D oscillator, $I_\rho$ and $I_\phi \equiv p_\phi$, is a
standard exercise \cite{36f}, and the Hamiltonian $H(\Omega =0)$ in terms
 of these  actions is
$$
H(\Omega =0) = 2I_\rho \omega_0 +|I_\phi|\omega_0.
\eqno(5.27)
$$
The term $\Omega p_\phi \equiv \Omega I_\phi$ in (5.26) will
not disturb the good actions, so that (5.26) becomes
 $$
  H =
2I_\rho \omega_0 +|I_\phi|\omega_0 -I_\phi \Omega \;\; .
\eqno(5.28)
$$

\vspace{2mm} \noindent
Application of (5.16) with $I=I_{\rho}$ and  $I=I_{\phi}$ then gives

$$
 \omega_\rho =2\omega_0 \; , \;\;
 \omega_\phi = \pm\omega_0 -\Omega \;\; ,
\eqno(5.29) $$

\vspace{2mm} \noindent
 where a positive (+) frequency in $\omega_{\phi}$ corresponds to a
counterclockwise rotation. Note that the
Foucault precession frequency $\omega_F \equiv -\Omega$ is true
for orbits of any shape, e.g. near linear (the usual case), elliptical, or
circular. The factor of 2 in $\omega_\rho$ in (5.29) is clear
intuitively: for $\Omega =0$, in one circuit of an  elliptical orbit, $\rho$
goes through two cycles  while $\phi$ goes through one.
Computing the phase of the pendulum $\phi(T_E)$ after one complete rotation
of the earth (period $T_E$) then gives
$$
\phi(T_E) = \pm \omega_0 T_E -\Omega T_E = \pm\omega_0 T_E -
2\pi\cos\theta \;\; .
\eqno(5.30)
$$
The term $\pm\omega_0 T_E$ is
the dynamical phase $\phi_D$ and $-2\pi\cos\theta$ is the
geometric Hannay angle $\Delta\phi_H$.

The examples in this section and the previous one led to a
nonvanishing Hannay angle $\Delta\theta_H$. It is easy to show
that the earlier example (particle in a 1D box with a slowly moving
wall) gives a vanishing result. The same is true for a particle in
a slowly rotating 1D box (fixed walls). Golin  \cite{36g} has  proved
that, for 1D systems with one adiabatic parameter $\lambda$, one
always has $\Delta\theta_H =0$ unless $\lambda$ undergoes a
rotational excursion through $2\pi$. However the proof does
not cover our rotating 1D box example (which has a hard wall
potential), since Golin \cite{36g} assumes a smooth potential. The case
of a slowly rotating 2D box of elliptical shape has been worked
out by Koiller \cite{36c}, and gives a nonvanishing $\Delta\theta_H$
depending in a complicated way on the area of the box.

\subsubsection{Nonintegrable Systems}

Nonintegrable systems have fewer than $f$ independent constants of the motion or good actions, where $f$ is the number 
of degrees of freedom.
For such systems, at least some of the possible motions are chaotic, and
for these at least some of the good action variables do not exist. Hence we have not been able
to proceed rigorously beyond the general relation (5.4) for chaotic motions.
(Despite the nonexistence of good actions $I$ for nonintegrable systems,
formal use of them and their adiabatic invariance have been used successfully in semiclassical
quantization schemes; see  sections  2.3, 3.1, 4.1, 4.3 and ref. \cite{6} for a method based on the Maupertuis Principles 
for conservative systems, and see refs. \cite{12, Skod} for a method based on adiabatic switching.)

As stated, we do not have rigorous arguments to draw consequences
from the UMP (5.4) for nonintegrable systems, but  some heuristic  arguments
can be given. For example, by choosing $H(\lambda(t))$ to vary slowly and linearly
with $t$, e.g. $H(\lambda(t)) =\lambda(t) H_0(q, p) =
(1+\dot\lambda t) H_0(q, p)$, where $\dot\lambda$ is small, we can \underline{derive} the adiabatic
invariance of the action $I$ for a 1D periodic system (earlier we
\underline{assumed}  that $I$ is an adiabatic invariant for such systems). 
Further, by
considering long trajectories of multidimensional systems, which start and end in some small region
in phase space, and which are not quite periodic in
general, we can formally show that, in a Poincar\'{e} cycle time or
in the limit $T\to\infty$ the long-path action $W$ is an
adiabatic invariant for quasiperiodic and chaotic trajectories of
multidimensional systems.  Ehrenfest \cite {Ehr} has previously shown
that $W$ is an adiabatic invariant for multidimensional  periodic
systems. For \underline{ergodic} systems, the phase volume $\Omega(E)$ is an
adiabatic invariant (see next paragraph), and since the adiabatic
invariant is apparently unique for such systems (Kasuga \cite{11}), it should be
possible to derive the adiabatic invariance of   $W$ from that
of  $\Omega$ for these systems, but we have been so far unable to do this \cite{note7}. We note that
counterparts of the Hannay angle have been established for
adiabatically evolving nonintegrable \cite{Mont} and ergodic \cite{Robb}
systems.

For one limiting case, complete chaos (ergodic motion) where the energy is the only constant of the motion,
and long trajectories typically cover essentially  the whole energy surface,
a rigorous adiabatic invariant has been found \cite {11,12,13}, i.e. the phase volume ${\Omega(E)}$, enclosed by the
energy surface $H(q,p) = E$:
$$
\Omega(E)=\int\ {dq}{dp} \Theta[ E - H(q,p)],
\eqno (5.31)
$$
where $\Theta$ is the step function, which is unity for positive argument. (For multidimensional systems, $dq dp$
stands for $ dq_1 dq_2 ... dq_f dp_1 dp_2 ... dp_f $.)

If $ H(q,p, \lambda(t))$ depends on a slowly varying parameter $\lambda(t)$ so that $E=E(\lambda(t))$,
then $\Omega(E(\lambda(t)))=\Omega(E(\lambda(0)))$. The adiabatic invariance of $\Omega(E)$ is intuitively plausible
\cite{Br}. First consider the initial ($t=0$) energy surface $H(q,p, \lambda(0)) = E(\lambda(0))$,
and its image under an \underline{arbitrary}
time development (i.e.not necessarily slow). The final (image) surface at the time $T$ is not in general a
constant energy surface, and hence not in general a dynamical surface. However, by Liouville's theorem the volume
enclosed by the final surface is equal to that enclosed by the initial surface. \underline {If}
the time-development from $t=0$ to $t=T$ \underline{is} sufficiently
slow then at each stage the surface which has evolved from the
$t=0$ surface \underline{is} a dynamical surface, by definition of adiabatic evolution,
where allowed motions transform into allowed motions \cite{Ehr}. In this case the final $t=T$
surface is a true constant energy dynamical surface, and hence $\Omega(E(\lambda(T)))=\Omega(E(\lambda(0)))$
from Liouville's theorem .

For ergodic systems, there are no good quantum numbers except for the energy-ordering quantum number
$N=1, 2, ...$, where $ E_1 < E_2 < ...$. A rough but useful semiclassical quantization scheme for ergodic systems
based on  $\Omega(E)$ is the Weyl rule \cite{13}

$$
\Omega(E_N)= N {h}^f,
\eqno (5.32)
$$

\vspace{2mm} \noindent
which implicitly defines $E_N$.
This is just a restatement of the early rule due to Planck \cite {Pl} that each quantum state occupies
a volume of size $h^f$ in
classical phase space. In practice, various improvements such as replacing $N$ by $(N+\alpha)$ as in (2.4)
are used \cite{13}. More subtle improvements can be obtained from Gutzwiller's trace formula \cite{9a}.
These relate fluctuations from the smooth dependence of $E$ on $N$ predicted by (5.32) to the classical periodic orbits.
In cases such as the $x^2y^2$ potential, where the classical  $\Omega(E)$ diverges \cite{Tom} due to the infinite
channels of the potential (see Fig.1), one can estimate the smooth part of $N(E) $ ( the cumulative number of states up to $E$)
quantum mechanically \cite{52} at low energy by an asymptotic expansion \cite{Tom}, with leading terms
$\varepsilon^{3/2}\log{\varepsilon}$ and $\varepsilon^{3/2}$,
where $\varepsilon = E(m^2/\hbar^4 C)^{1/3}$ is dimensionless. Application of the generalization
 of (5.32), i.e. $N(E_N)=N$, then gives reasonable
estimates of the low-lying state energies \cite {Whe, Tom}.

\subsection{The Classical Hellmann - Feynman Theorem}

 Another consequence of the UMP (5.4) is the classical version of the Hellmann - Feynman theorem \cite{Kin}

$$
\frac{\partial E(I, \lambda)}{\partial \lambda}= <\frac{\partial H(q,p,\lambda)}{\partial \lambda}>_I.
\eqno(5.33)
$$

\vspace{2mm} \noindent
Here $H(q,p,\lambda)$ depends on a parameter $\lambda$, so that the energy $E(I, \lambda)$ depends on $\lambda$,
as well as on the (constant) action $I$, and $<...>_I$ denotes a time-average over a period $T(I)$ at fixed action $I$.
We assume a one-dimensional periodic system for simplicity.

The quantum version of the theorem \cite{Hell} is nowadays much better known, i.e.
$$
\frac{\partial E_n(\lambda)}{\partial \lambda}= <\frac{\partial\hat H(\lambda)}{\partial \lambda}>_n,
\eqno(5.34)
$$
where $<...>_n= <n|...|n>$, and where the Hamiltonian operator $\hat H(\lambda)$
depends on some parameter $\lambda$, so that the energy eigenvalue $E_n(\lambda)$
and eigenfunction $\psi_n(\lambda)$ do also, because of
$$
\hat H(\lambda)\psi_n(\lambda) = E_n(\lambda)\psi_n(\lambda).
\eqno(5.35)
$$
Proof of (5.34) is simple; we apply $\partial/\partial\lambda$ to
both sides of the relation
$E_n(\lambda) =
 \langle\psi_n(\lambda)| \hat H(\lambda)|\psi_n(\lambda)\rangle$,
and note that the two terms
involving $\partial\psi_n(\lambda)/\partial\lambda$ add to zero
because of (5.35), the hermiticity of $\hat H(\lambda)$, and the fact that
 we assume $\langle\psi_n(\lambda)|\psi_n(\lambda)\rangle = 1$
for all $\lambda$.

It is clear that (5.33) is the classical limit of (5.34), since as
$\hbar \to 0$ (or $n\to\infty$) $\langle ...\rangle_n \equiv
\langle n|...|n\rangle$ approaches a time average over a period,
and fixing $n$ corresponds to fixing $I$, since $I \simeq n\hbar$. To derive (5.33) from (5.4) we consider $H(\lambda(t))$ to evolve adiabatically, so that $\lambda$ and $T$ in (5.4) can be related. The derivation is intricate, however, and it is simpler to proceed instead  from the RMP (2.3) for conservative systems. We apply (2.3) to two periodic trajectories of $H(\lambda)$, the true one with period $T(\lambda)$, and a virtual one taken as the true trajectory of $H(\lambda+\Delta \lambda)$ with period $T(\lambda+\Delta \lambda)$ . Since first-order variations in
 $\bar E \equiv  <H(\lambda)>$ vanish (at fixed $W = 2\pi I$), we have

$$
<H(\lambda)>_{\lambda+\Delta \lambda} = <H(\lambda)>_{\lambda}+ O(\Delta\lambda^2),
\eqno(5.36)
$$

\vspace{2mm} \noindent where $<...>_{\lambda}$ is an average over the trajectory with period $T(\lambda)$, and it is understood that $I$ is the same for both $<...>_{\lambda}$ and $<...>_{\lambda+\Delta \lambda}$. From (5.36) we get for $\Delta\bar E \equiv <H(\lambda + \Delta\lambda)>_{\lambda+\Delta \lambda} - <H(\lambda)>_{\lambda}$ to $O(\Delta\lambda)$ accuracy
$$
\Delta\bar E = <H(\lambda + \Delta\lambda)>_{\lambda+\Delta \lambda} - <H(\lambda)>_{\lambda+\Delta \lambda} = 
<H(\lambda + \Delta\lambda)- H(\lambda)>_{\lambda + \Delta\lambda}.
$$
Again neglecting $O(\Delta\lambda^2)$ terms we have
$$
\Delta\bar E = <H(\lambda + \Delta\lambda)- H(\lambda)>_{\lambda}.
\eqno(5.37)
$$
 Eq(5.37) yields $\partial{\bar E}/\partial \lambda = < \partial H(\lambda) /\partial\lambda>_{\lambda}$, and since $\bar E = E$ for the true trajectory we get  
$\partial{E}/\partial \lambda = < \partial H(\lambda) /\partial\lambda>_{\lambda}$, the Hellmann-Feynman relation (5.33) apart from notation.

As a simple example of (5.33) consider a 1D harmonic oscillator,
with Hamiltonian $H(x,p) = p^2/2m + k x^2/2$ and $E(I) =
I\omega$, where $\omega = (k/m)^{1/2}$ is the angular frequency.
Choose $\lambda = k$, the force constant. The left side of (5.33)
is $E/2k$ and the right side is $<x^2/2> = <V>/k$. Thus we have
$<V> = E/2$, as is well known from the virial theorem \cite {1b}. If we
choose $\lambda =m$, we find similarly $<K> = E/2$.

As a second example, consider the bound motion of a 1D system with
Hamiltonian $H(x,p) = p^2/2m +V(x)$. We change variables $x, p
\to x', p'$ via the transformation $x' = \lambda x$, $p' =
p/\lambda$, where $\lambda$ is an arbitrary scale factor. This
transformation is canonical and preserves the Hamiltonian \cite {39a},
i.e. $H(x, p) = H_\lambda(x', p')$, where
$$
H_\lambda(x', p') = \lambda^2 \frac{p'^2}{2m} +
V(\frac{x'}{\lambda}) \;\;  .
\eqno(5.38)
$$
The energy of a given state of motion is unchanged by the change
 from $H$ to $H_\lambda$ for any value of $\lambda$. We now use (5.38) in the Hellmann-Feynman
 relation (5.33). The left hand side vanishes. The right hand
 side derivative is
 $$
 \frac{\partial H_\lambda}{\partial\lambda} = \frac{\lambda
 p'^2}{m} - \frac{x'}{\lambda^2} V' = \frac{2}{\lambda} K -
 \frac{x}{\lambda}V',
 $$
 where $V' = \partial V(x)/\partial x$, and we have reverted  to
 the original variables in the second form. Hence we get
 $$
 <2K> = <x V'> \;\; ,
 \eqno(5.39)
 $$
 i.e. the virial theorem for an arbitrary potential $V(x)$.

\section{Variational Principles in Relativistic Mechanics}

Hamilton's  Principle is widely used in Classical Relativistic Mechanics and in the Classical Theory of
Electromagnetic and Gravitational Fields to derive covariant equations of motion (see, e.g.\cite{14}).
As for Maupertuis' Principle, it is widely believed that it is not well suited for that purpose
because when operating with the energy one loses explicit covariance. In this section we demonstrate,
first, how to use Maupertuis' Principle for the derivation of the covariant equations of motion and,
the second, how to use it for the solution of specific problems.

Consider a relativistic particle with mass $m$ and charge $e$ moving in an arbitrary external electromagnetic
field, described by a four-potential with contravariant components $A^{\alpha}=(A_0,A_i)=(\phi,A_i)$,
and covariant components $A_{\alpha}=(A_0,-A_i)=(\phi,-A_i)$, where  $\phi$ and  $A_i$  (for $i=1,2,3$) are
the usual scalar and vector potentials. Hamilton's Action for this system can be written in the Lorentz
invariant form \cite{14, Lan}

$$
S = m\int\ ds + e \int\ A_{\alpha} {dx^{\alpha}},
\eqno(6.1)
$$
where we use the sign of Lanczos \cite{Lan} for $S$, opposite to that of Landau and Lifshitz \cite{14}.
The sign can be chosen arbitrarily; our choice allows us to use below the orthodox definitions
$p = {\partial L}/{\partial{v}}$ and $H=pv - L$.
 The four-dimensional path runs from the initial point $x_A = (x^0_A, x^1_A, x^2_A, x^3_A)$ to the final point $x_B = (x^0_B, x^1_B, x^2_B, x^3_B)$
 in four-dimensional space-time with  corresponding proper times $s_A$ and $s_B$.
Here $ds$ is the infinitesimal interval of the path (or of the proper time)
$ds^2 = dx_{\alpha}dx^{\alpha} = g_{\alpha\beta}dx^{\alpha}dx^{\beta}= dx_0^2  - dx_i^2=dt^2  - dx_i^2$,
the metric has signature $(+,-,-,-)$ and we use
the summation convention and units
with $c=1$, where $c$ is the velocity of light. $S$ is itself not gauge invariant, but a gauge transformation
$ A_{\alpha}\rightarrow A_{\alpha} + {\partial f}/{\partial x^{\alpha}}$ ( for arbitrary $f$) adds
only constant boundary points terms to $S$, so that $\delta S$ is unchanged.
The HP is thus gauge invariant.

If we introduce a parameter $\tau$ along the four-dimensional path (proper time $s$ along the true
 or any particular virtual path are valid choices) the action $S$ can be rewritten in the
form

$$
S = m\int\ [ v_{\alpha}v^{\alpha}]^{1/2} {d\tau} + e \int\ A_{\alpha}v^{\alpha}{d\tau},
\eqno(6.2)
$$
where $v^{\alpha}={d x^{\alpha}}/{d\tau}$. In general, $\tau$ is both frame-independent and path-independent. A path-independent 
parameter is here invaluable for variational purposes, since we then do not have to vary the parameter when we vary the path. 
After the variations have been performed, we can then choose $\tau = s$, etc.
The limits of the integrals are the invariants
$\tau_A$ and $\tau_B$. With respect to parameter $\tau$ we get the covariant Lagrangian

$$
L = m[ v_{\alpha}v^{\alpha}]^{1/2} + e v_{\alpha}A^{\alpha}  ,
\eqno(6.3)
$$
and corresponding conjugate momenta
$$
p^{\alpha} = \frac{\partial L}{\partial v_{\alpha}}= m v^{\alpha}/ [ v_{\beta}v^{\beta}]^{1/2} + eA^{\alpha},
\eqno(6.4)
$$
and Hamiltonian
$$
H = p_{\alpha}v^{\alpha} - L = 0.
\eqno(6.5)
$$

Thus in this particular covariant treatment the Hamiltonian is trivial. The 
dynamics is hidden in a constraint.
One can see that the four conjugate momenta are not independent variables but satisfy the constraint
$$
(p_{\alpha} - e A_{\alpha}) (p^{\alpha} - e A^{\alpha})= m^2 ,
\eqno(6.6)
$$
which is obvious from (6.4).
This constraint can be taken into account with the help of the Lagrange multiplier method,
i.e. with the help of an effective Hamiltonian \cite{Lan,Dir}
$$
H_{eff} = {\lambda} [(p_{\alpha} - e A_{\alpha})(p^{\alpha} - e A^{\alpha})- m^2 ].
\eqno(6.7)
$$
We easily find that the Hamilton equations of motion with $H_{eff}$  are equivalent to
$$
{\lambda}= 1/2m
$$
(for ${\tau}= s$, the true trajectory proper time ) and to the covariant Lorentz equations of motion
for a charged particle in an external field:
$$
m \frac {d v_{\alpha}}{d s}= e F_{\alpha\beta} v^{\beta},
$$
$$
F_{\alpha\beta}= \frac{\partial A_{\beta}}{\partial x^{\alpha}} - \frac{\partial A_{\alpha}}{\partial x^{\beta}},
\eqno(6.8)
$$
i.e. the same equations as are found from the Lagrangian (6.3) (see, e.g. \cite{14}).

On the other hand  the Hamilton equations with $H_{eff}$ are equivalent to the MP or to the RMP with the same $H_{eff}$ .
Thus one can use any of the four Variational Principles (HP, RHP, GMP, RMP) either with Lagrangian (6.3) or with effective Hamiltonian (6.7)
for the derivation of the covariant equations of motion. It is to be noted that in the covariant formulation of relativistic
mechanics, the Hamiltonian (a Lorentz scalar) is not equal to the energy ( the time component of a four-vector).
Thus $\bar E$ in the Maupertuis Principle must be interpreted as
$$
 \bar E\equiv \bar H_{eff} =\frac {1}{\triangle\tau}\int\limits_{\tau_A}^{\tau_B}\ {d\tau} H_{eff},
\eqno(6.9)
$$
where $\triangle\tau=\tau_B-\tau_A$. By contrast, in the noncovariant formulation discussed below,
the Hamiltonian is equal to the energy $K+V+m$ .

There are other, related, formal difficulties arising from (6.5). This condition ($H=0$) is due to the fact that the
Lagrangian $L$ in (6.3) is homogeneous and of first degree in $v_{\alpha}$ (i.e. essentially linear in
velocities $v_{\alpha}$), so that $v_{\alpha}\partial L/\partial v_{\alpha} = L $ and hence the difference
$v_{\alpha}\partial L/\partial v_{\alpha} - L $
(the Hamiltonian) vanishes \cite{98a}.
  Since $H=0$, the Legendre transform relation (6.5) between $L$ and $H$ simplifies to
$$
L = p_{\alpha}v^{\alpha},
\eqno(6.10)
$$
and hence the Legendre transform relation between $W$ and $S$, i.e. $W-S=\bar H \Delta\tau$  (where ${\bar H} \equiv {\bar E}$), simplifies to 

$$
S =W=\int\limits_{\tau_A}^{\tau_B}  p_{\alpha}v^{\alpha} {d\tau} =  \int\limits_{x_A}^{x_B}  p_{\alpha} dx^{\alpha}.
\eqno(6.11)
$$
Because $H\equiv 0$, direct application of the GMP  ($ \delta W)_{\bar E=0} = 0$
would be cumbersome, and direct
application of the  RMP ($\delta \bar E)_{W}=0$ would be impossible.

All these difficulties can be traced to the well studied difficulty \cite{1b, Lan, Doug, Bar,  For}
of applying an action principle with
 kinematic \cite{Non} nonholonomic
constraints, such as (6.6) for the momenta, or for the velocities $v_{\alpha} v^{\alpha} = 1$
for proper time $s$, or
$v_{\alpha} v^{\alpha} = ({ds}/{d\tau})^2$ for arbitrary parameter $\tau$.

 One way out of the difficulties is to introduce the constraint with a Lagrange multiplier for
 either the Lagrangian or
 Hamiltonian (as is done above for $H_{eff}$). Another way out is to exploit the nonuniqueness of the Lagrangian. For example
 \cite {Doug}, the covariant Lagrangian $L$ (6.3) is equivalent to the covariant Lagrangian $L^{'}$, where (with $\tau$ as parameter)

$$
L' = \lambda(\tau)\frac{m}{2} v_{\alpha}v^{\alpha} + e v_{\alpha}A^{\alpha},
\eqno(6.12)
$$
where $\lambda(\tau)= {d\tau}/{ds}$. It is easily verified that (6.12) generates
the same Euler-Lagrange equations of motion as (6.3). For $\tau = s$ (the  proper time of the path in question), (6.12) simplifies to

$$
L' = \frac{m}{2} v_{\alpha}v^{\alpha} + e v_{\alpha}A^{\alpha},
\eqno(6.13)
$$
which is often employed \cite {1b,Bar}. The advantage of (6.12) and
(6.13) is that they are nonhomogeneous in $v_{\alpha}$,
and hence will generate a nontrivial corresponding Hamiltonian $H'$. For (6.13), for example, we find

$$
H' = \frac{1}{2m} (p_{\alpha} - e A_{\alpha})(p^{\alpha} - e A^{\alpha}).
\eqno(6.14)
$$

\vspace{2mm} \noindent Note that $H'$ (6.14) and $H_{eff}$ (6.7) (for $\tau=s$)  differ only by a constant.
Use of the covariant $L'$ and $H'$ in place of $L$ and $H$ removes the difficulties of using
all four variational principles (HP, RHP, GMP, RMP).  For $L$ and $H$, only the first
two can be used. Note again that $\bar E$ is computed from $H'$ as in (6.9), with $H'$ in place of
$H_{eff}$.

  Now we come to
the question of the solution of specific physical problems. In this case explicit covariance
usually is not of great importance and it is more useful to find an appropriate convenient frame.
For any given frame we can choose $\tau=t$ (the coordinate time) and obtain  the complete set of nontrivial quantities

$$
L = -m(1-v^2)^{1/2} - e \phi + e{\bf v \cdot A}, \;\;\; S=\int\limits_0^{T} {L}{dt},
$$

$$
{\bf p} = \frac {m\bf v}{(1 - v^2)^{1/2}} + e{\bf A},
$$
$$
W = \int\limits_0^{T} {\bf p\cdot v} {dt} =  \int\limits_0^{T} {dt}[\frac{ mv^2}{(1-v^2)^{1/2}} + e{\bf v\cdot A}],
\eqno(6.15)
$$
\vspace{3mm}
$$
H = \frac{ m}{(1-v^2)^{1/2}} + e \phi, \;\;\;  \bar E = \frac{1}{T} \int^T_0 H dt.
$$

\vspace{2mm} \noindent The expression for $L$ in (6.15) is derived from (6.3) or (6.12) by choosing $\tau=t$,
except that we have used the freedom to change the sign of $L$ to agree
with the standard choice in the noncovariant case \cite{1b}. Again we have chosen
 $t_A = 0$ and $t_B = T$. Note that the
nonrelativistic form $W = \int^T_0 2K dt$ for the Maupertius action is not valid
here, and also note that the Hamiltonian $H$ is equal to the energy
$K +e\phi+m$, despite the fact that $L \neq K-V$.

As a result the complete set of four variational principles is available in any particular frame.
It is matter of taste or practical convenience to use one principle or another.
From a practical point of view there is no difference in the usage of variational principles in nonrelativistic mechanics
and in relativistic mechanics in a given frame.

As an example of the usage of the GMP for relativistic mechanics consider the motion of an electron in a uniform magnetic
field ${\bf B} = (0,0,B)$. A corresponding vector potential in the Coulomb gauge (${\bf \nabla}\cdot {\bf A} = 0$) is
${\bf A} ={\bf B}\times{\bf r}/2$, or ${\bf A} = (-yB/2, xB/2, 0)$ \cite{106a}.
The solution of (6.8) in this case is well known - the electron moves with constant velocity
along the direction of the
field (along the z-axis in our case) and along a circular orbit in the $xy$ plane with the cyclotron frequency
$$
\omega = \frac{|e|B}{E},
\eqno(6.16)
$$
where $E$ is the total energy of the particle including the kinetic energy of the motion in the z -direction.

To calculate this frequency from the GMP we take as a trial trajectory
$$
 x= A\cos{\omega}t, \;\;\; y =  A \sin{\omega}t,\;\;\; v_z=0,
$$
where $\omega$ and $A$ are free parameters. After some simple algebra we get

$$
\bar E  = \frac{m}{(1-\omega^2 A^2)^{1/2}},
$$
$$
W = {2\pi}A^2  [ \frac{m\omega}{(1-\omega^2 A^2)^{1/2}} -\frac{|e|}{2}B ].
\eqno(6.17)
$$

\vspace{2mm} \noindent In this case it seems that an economical method is to use the GMP (2.2).
For this purpose one has to rewrite $W$ as a function of $\bar E$
$$
W = \frac{2\pi}{\omega^2}  [ 1- (\frac{m}{\bar E})^2 ][ \omega \bar E - \frac{|e|}{2}B],
\eqno(6.18)
$$
and to calculate the variation    $\delta{W}$ at fixed  $\bar E$. The GMP (2.2)
translates here to $({\partial W}/{\partial \omega})_{\bar E}=0$.
One then easily finds the  analytical solution

$$
\omega = \frac{|e|B}{\bar E}.
$$
The exact solution corresponds to the case when  $\bar E = E$. For our trial
trajectory, we happen to have $\bar E = E$ and the approximate solution
coincides with the exact one ( similar to the case of a harmonic oscillator).

This simple example demonstrates how to use variational principles for relativistic systems.

\section{Classical Limit of Quantum Variational Principles}

 In section 2.1 we mentioned that the classical limit of the
Schr\"{o}dinger Quantum Variational Principle
is the Reciprocal Maupertuis Principle for periodic or other steady-state motions, i.e.
$$
\left(\delta\frac{<\psi|\hat H|\psi>}{<\psi|\psi>}\right)_n =0
\stackrel{n\gg 1}{\longrightarrow} \left(\delta \frac{1}{T} \int^T
_0 H(q, p)dt \right)_W = 0 \;\; ,
\eqno (7.1)
$$
 where $\hat H$ is the Hamiltonian operator corresponding to the classical 
Hamiltonian $H(q,p)$.
 Here $T$ is the period for periodic motions, or $T \to \infty$ for other steady-state motions (quasiperiodic or chaotic).
Solution of the left-hand side yields \cite{81a} the stationary state $\psi_n$ with the
energy eigenvalue $E_n$, i.e.
$$
\hat H \psi_n = E_n \psi_n ,
\eqno (7.2)
$$
the time-independent  Schr\"{o}dinger equation. Details of the proof of (7.1) (which is
straightforward only for integrable systems) are given in \cite{5}. In ref.\cite{GKN3} it is explained how
Schr\"{o}dinger, in his first paper on wave mechanics in 1926, in essence worked
backwards, i.e. starting from the right hand side of  (7.1),  he ``derived" the left hand side, and thereby discovered wave
mechanics. In his second paper on wave mechanics in 1926 he retracted this argument
because his version of the right hand side of (7.1) was not quite correct. In this second
paper he presented an alternative argument as a basis for wave mechanics (now in the standard textbooks) based on the analogy
between geometric and wave optics on the one hand, and particle and wave mechanics on the other.

We consider now the generalization of (7.1)  to nonstationary states $\psi(t)$,
where the true dynamical states satisfy the time-dependent Schr\"{o}dinger equation
$$
 i\hbar  \partial_t \psi(t) = \hat H \psi(t),
\eqno (7.3)
$$

\vspace{2mm} \noindent where $\partial_t = {\partial} /{\partial t}$, and the initial state $\psi(0)$ is given.
 For notational simplicity
we consider a single particle in three dimensions, so that $ \psi= \psi({\bf r}, t)$, and

$$
\hat H  = K + V = - \frac{\hbar^2}{2m}{\bf \nabla}^2 + V(\bf{r}) ,
\eqno (7.4)
$$
where $K=  {\bf p} ^2 /{2m}$ and $\bf{p}$ = $(\hbar/i)\bf{\nabla}$. The arguments which follow are also
valid for nonconservative systems, where $\hat H=\hat H(t)$.

We  shall consider here the quantum time-dependent variational principle due to Frenkel and Dirac \cite{22,115b, 115a}
for the wave function $\psi(t)$, but we note the existence of other quantum time-dependent variational principles,
e.g. for the time-evolution operator and transition amplitudes \cite{23},
 for the Heisenberg operator version  \cite{24}  and for the matrix mechanics version \cite{Green}
 of the Hamilton Principle, for the density matrix \cite{Ebo}, for the mean value of an
arbitrary observable  \cite{88a}, and for time-dependent density functional theory \cite{Run}.
( For completeness, we note that there are also other \cite{123a} quantum time-independent variational principles, 
in addition to that of Schr\"{o}dinger.)

 The original Frenkel-Dirac variational principle is \cite{22}

$$ <\delta\psi|\hat H - i\hbar\partial_t|\psi>=0,
\eqno (7.5)
$$

\vspace{2mm} \noindent
for arbitrary variations $\delta\psi({\bf r},t)$ around the true
dynamical state $\psi({\bf r},t)$. Since (7.5) is true for all
possible variations $\delta\psi$, then $(\hat H -
i\hbar\partial_t)\psi=0$ for the true states, as in (7.3). By
adding (7.5) and its complex conjugate and using the hermiticity of
$\hat H$ we can transform (7.5) to \cite{25}
$$
\delta <\psi|\hat H -i\hbar\partial_t|\psi> = -i\hbar \partial_t <\psi |\delta\psi >.
\eqno (7.6)
$$

Integrating over an arbitrary but fixed time interval $(0, T)$, and assuming
$\psi(t=0)$ is fixed, we find \cite{26} from (7.6)
$$
\delta\int_0^T <\psi |\hat H -i\hbar \partial_t |\psi > dt =
-i\hbar <\psi(T)|\delta\psi(T)>.
\eqno (7.7)
$$

Note that (7.7) is not in the traditional form for a variational
principle $\delta\int Ldt =0$, because of the right hand side.
It is more like the unconstrained principles discussed earlier.
Some authors drop the right hand side without comment \cite{27},
but it is required if arbitrary variations $\delta\psi$ in $\psi$ are
allowed; otherwise (7.7) will not generate the Schr\"{o}dinger equation (7.3) and its complex
conjugate \cite{25}. It is possible to obtain a vanishing
right-hand side by restricting the variations $\delta\psi$ in
various ways \cite{29}, e.g. requiring the trial wave functions $\psi(t)$ to be
normalized at all times. Another way is to write the right hand side as
$-i\hbar <\delta\psi(T)|\psi(T)>^{*}$, formally regard $\psi$  and its complex conjugate $\psi^{*}$ as
independent (see below), and assume $\psi^{*}(T)$ (as well as $\psi(0)$)
is fixed. However, we find it simplest to retain the general form (7.7), with
no constraints on $\psi(t)$ except for the fixing of the initial condition $\psi(0)$.
(When dealing explicitly with bound states later, we assume boundary conditions on
$\psi({\bf r}, t)$.)

By writing out the scalar products $<A|B> \equiv\int d{\bf r} A({\bf r})^{*} B({\bf r})$ in (7.7)
and using the hermiticity of ${\bf p} = (\hbar/i)\bf\nabla$, we can write (7.7) in  field
theory form
$$
\delta\int_0^T dt \int d{\bf r} {\cal L} = -i\hbar\int d{\bf r} \psi(T)^{*}
\delta\psi(T) \;\; ,
\eqno (7.8)
$$
where the Lagrangian density ${\cal L}$ is
$$
{\cal L} (\psi, \psi^*) =\frac{\hbar^2}{2m}|{\bf\nabla}
\psi|^2 + V|\psi|^2  -i\hbar \psi^* \partial_t \psi \;\; .
\eqno (7.9)
$$
Since $\psi$ is a complex function, whose real and imaginary parts can be
varied independently, it is well known \cite{30} that $\psi$ and
$\psi^*$ can be regarded as independent for variational purposes.
These two variations in (7.8) with ${\cal L}$ given by (7.9) will
generate the Schr\"{o}dinger equation and its complex conjugate.

Following Feynman \cite{27}, we change variables $\psi, \psi^* \to
A, S$ (or $\rho, S$) in the Lagrangian density ${\cal L}(\psi, \psi^*)$, 
where $A({\bf r}, t)$ is the amplitude and
$S({\bf r}, t)$ the phase of the wave function, via the Madelung
transformation \cite{31}:
$$
\psi = Ae^{iS/\hbar} = \sqrt{\rho} e^{iS/\hbar} \;\; ,
\eqno (7.10)
$$
where $\rho = A^2 = |\psi|^2$ is the probability density.
In terms of $\rho$ and $S$ we have
$$
{\cal L}(\rho, S) = \rho\frac{({\bf \nabla} S)^2}{2m} +
\frac{\hbar^2}{2m}({\bf \nabla} \sqrt{\rho})^2 +\rho V +
\rho\partial_t S -\frac{i\hbar}{2} \partial_t \rho.
\eqno (7.11)
$$
In terms of $\rho, S$ the boundary term in (7.8) is
$$
<\psi(T)|\delta\psi(T)> = \int d{\bf r} \sqrt{\rho(T)}
\delta\sqrt{\rho(T)} + \frac{i}{\hbar}\int d{\bf r} \rho(T)
\delta S(T) \;\; .
\eqno (7.12)
$$

We now take the classical ($\hbar \to 0$) limit of (7.8). Using
(7.11) and (7.12) we get
$$
\delta\int_0^T dt \int d{\bf r} \rho[\frac{({\bf\nabla} S)^2}{2m}
+V +\partial_t S] = \int d{\bf r} \rho(T)\delta S(T) \;\; .
\eqno (7.13)
$$

\vspace{2mm} \noindent
We have assumed that in the limit $\hbar \to 0$, $\rho$ and $S$ approach well defined limits
( which we continue to denote by $\rho$ and $S$). In particular we assume that $\nabla\sqrt{\rho}$
does not contain an $O(\hbar^{-1})$ term \cite{Hol}.
We vary $\rho$ and $S$ independently in (7.13). Varying $\rho$, with $S$
fixed, gives
$$
\int\int dtd{\bf r} \delta\rho [\frac{({\bf\nabla} S)^2}{2m} +V
+\partial_t S] =0.
$$
Since $\delta \rho({\bf r}, t)$ is arbitrary, the bracket [...] must
vanish, so that
$$
\partial_t S +\frac{({\bf \nabla} S)^2}{2m} + V =0 \;\; .
\eqno (7.14)
$$
This is the classical Hamilton-Jacobi equation \cite{32} for the classical action
$S({\bf r}, t)=\int_{{\bf r}_0, 0}^{{\bf r},t} Ldt'$ along true paths.

Next we vary $S$ in (7.13) with $\rho$ fixed. We integrate the $\delta({\bf\nabla} S)^2$
term by parts using
$$
\int d{\bf r} \rho({\bf\nabla} \delta S) \cdot {\bf\nabla} S = -\int
d{\bf r} \delta S {\bf\nabla} \cdot (\rho{\bf\nabla} S) \;\; ,
\eqno (7.15)
$$

\vspace{2mm} \noindent
where we drop a surface term by assuming $\rho \to 0$ for
${\bf r} \to\infty$. This assumes the corresponding quantum state
has finite norm. We integrate the $\delta(\rho\partial_t S)$ term
by parts using
$$
\int_0^T dt\rho\partial_t \delta S = \rho(T)\delta S(T) -
\int_0^T dt\delta S \partial_t \rho \;\; ,
\eqno (7.16)
$$
where we use $\delta S(t=0) =0$ which follows from the fixing of
$\psi(t=0)$. Using (7.15) and (7.16) in (7.13) thus gives
$$
\int_0^T dt \int d{\bf r} [{\bf\nabla} \cdot(\rho\frac{{\bf\nabla} S}{m})
+\partial_t \rho]\delta S =0.
$$
Since this is true for arbitrary $\delta S({\bf r}, t)$, we get
$$
\partial_t \rho +{\bf\nabla} \cdot(\rho\frac{{\bf\nabla} S}{m}) =0 \;\; .
\eqno (7.17)
$$
This is the continuity equation, reflecting conservation of total
probability, since \cite{32} ${\bf\nabla} S = {\bf p} = m\bf v$ is
the momentum of the particle. Note that if the right hand side of
(7.13) were missing, we would not get the correct continuity
equation. In contrast to what we have done here, Feynman \cite{27} keeps the
$\hbar$ terms in (7.11), and the variational principle then leads to quantum
generalizations \cite{130a} of the Hamilton - Jacobi (7.14) and continuity
(7.17) equations.

Thus we see that the classical limit of the quantum time-dependent
variational principle is the classical variational principle for
the Hamilton-Jacobi and continuity equations. Interestingly the
classical limit of the quantum time-dependent variational principle
leads to a classical variational principle for the Hamilton action
$S$ along the true trajectories, whereas the classical limit of the quantum
time-independent variational principle in the form of the left side of (7.1)
leads to a classical variational principle  for
the trajectories themselves (RMP -the right side of (7.1)). An alternative form \cite{95a} of the quantum
time-independent variational principle,

$$ \delta<\psi|\hat H - E|\psi>=0,
\eqno (7.18)
$$

\vspace{2mm} \noindent
will generate in the classical limit a variational principle for the
Maupertuis (or time-independent) action $W(\bf r)$ along the true paths. In (7.18) $E$ is a constant
Lagrange multiplier introduced to remove the normalization constraint $<\psi|\psi>=1$.
The extremizing $\psi(\bf r)$'s which satisfy (7.18) obey
$$ (\hat H - E)\psi=0,
\eqno (7.19)$$
so that $E$ is the energy eigenvalue corresponding to the eigenfunction $\psi$.

To derive the classical limit of (7.18) we proceed much as before. (We assume $\hat H$
is time-independent in this section.) We first apply the Madelung transformation
to $\psi(\bf r)$, i.e.
$$
\psi ({\bf r})= A({\bf r})e^{iW({\bf r})/\hbar} \;\; ,
\eqno (7.20)
$$
where $A({\bf r})$ is the amplitude and  $W({\bf r})$ the phase. Following similar steps
as above, we find (7.18) becomes in the classical ($\hbar \to 0$) limit
$$
\delta\int d{\bf r} \rho({\bf r})[\frac{({\bf\nabla} W)^2}{2m}
+V - E] = 0 \;\; ,
\eqno (7.21)
$$
where $\rho({\bf r})= A({\bf r})^2= |\psi({\bf r})|^2$.

Varying $\rho({\bf r})$ in (7.21) with $W({\bf r})$  fixed gives

$$
 \frac{({\bf\nabla} W)^2}{2m}+ V - E = 0 \;\; ,
\eqno (7.22)
$$

\vspace{2mm} \noindent
the time-independent Hamilton-Jacobi equation \cite{32} for the
Maupertuis action $W({\bf r}) = \int_{{\bf r}_0}^{\bf r} {\bf p}\cdot d{\bf r}^{'}$
 along true paths. If we vary $W({\bf r})$ in (7.21)
with fixed $\rho({\bf r})$ we get the continuity equation
$$
{\bf\nabla} \cdot(\rho\frac{{\bf\nabla} W}{m})=0,
\eqno (7.23)
$$
since \cite{32} ${\bf\nabla} W = {\bf p} = m{\bf v}$ is the momentum of the particle.

In ref.\cite{5}, we show how the classical limit of (7.18)
can be taken in a different way (using standard 1D WKB wave functions) and for a 1D periodic
orbit leads to $ \delta (S+ET)=0$, i.e. the UHP for a periodic orbit of period $T$.

We note that, just as for the right hand side of (7.1), the derivation of the classical limits (7.14) and
(7.22) is somewhat formal for the chaotic motions of nonintegrable systems, since 
classical values of $S$ and $W$ may not exist as well defined functions for such systems. Unlike the
nonrelativistic limit ($v/c \to 0$), which is regular, the limit $\hbar /S \to 0$ or $ \hbar /W\to 0$
is singular, which makes the classical limit subtle and interesting \cite{Ber}.

\section{Summary and Conclusions}

We summarize the two groups of classical  variational principles reviewed in this paper. First we have the Maupertuis principles: 
$$
\delta{W} - {T}\delta{\bar{E}} = 0,\;\; (\delta W)_{\bar E} = 0,\;\; (\delta \bar {E})_W =0.
\eqno(8.1)
$$
The first of these is the unconstrained form (UMP), the second the general Maupertuis principle (GMP), and the third is the reciprocal 
Maupertuis principle (RMP). By Legendre transformation of the principles in (8.1) we obtain the Hamilton principles:
$$
\delta{S} + {E}\delta{T} = 0,\;\;(\delta S)_{T} = 0,\;\; (\delta T)_{S} = 0 .
\eqno(8.2)
$$ 
The first is the unconstrained form (UHP), the second the Hamilton principle (HP), and the third the reciprocal Hamilton principle (RHP).
In all these principles, the end-positions of the trajectory $q_A$ and $q_B$ are held fixed, and the notation indicates what else is 
 fixed and what is  varied.

We have given a number of examples showing how to use these principles to solve practical problems, focussing particularly on the 
three Maupertuis principles. Both classical  and  semiclassical applications have been given, and some comparisons are made with 
the results from quantum variational principle calculations. The classical limits of two quantum variational principles (time-dependent 
and time-independent) are given. In particular, the RMP is the classical limit of the Schr\"{o}dinger time-independent quantum variational 
principle. The RMP is thereby naturally adapted to semiclassical applications

In the applications, the variational principles ( both classical and quantum) are solved by the direct   variational method, where one 
guesses a trial solution containing adjustable parameters, and adjusts the parameters to satisfy the variational principle. 
The Euler-Lagrange differential equation is not required in this method. When we choose a simple trial solution, the results can 
be found simply and analytically.

 Using the direct variational method the RMP and UMP have been used to estimate semiclassically the energy 
levels in several examples. The levels $E_{n_1, n_2}, ...$ can be expressed in analytical form, and the accuracy is quite acceptable in most cases, even for nonintegrable systems which are usually treated by much more complicated semiclassical methods, e.g. the Gutzwiller trace formula \cite {9a}, or quantizing the Birkhoff-Gustavson normal form \cite{x}. It would be useful to have comparisons of the results from the Maupertuis Principle-based method with those from the latter two and other methods for a number of specific problems. The methods should be compared for accuracy of results and computational complexity. Some limited comparisons of the Maupertuis Principle and normal form methods have been done for the $x^2 y^2$ oscillator of sec. 4.1 in refs. \cite{42l, 42q}, and limited comparisons for this model of results from the Maupertuis Principle and the semiclassical adiabatic methods \cite {9} are presented in Table 1 of sec. 4.1.1, but much more extensive comparisons would be of interest.

Variational principles appeal to various individuals for differing reasons, both aesthetic \cite {y} 
 and practical. We have stressed the utility for classical and semiclassical particle dynamics calculations in this review. 
Engineers  too make heavy use of variational principles in solving practical problems of
classical continuum mechanics \cite {z}. On rare occasions variational principles have even led to new laws of physics \cite {z1}.

\section{Acknowledgements}

We gratefully acknowledge the Natural Sciences and Engineering Research Council of Canada for
financial support of this work. We thank Donald Sprung for a careful reading of the manuscript, David Garfinkle for 
helpful remarks on relativistic mechanics and Stepan Bulanov for help with TeX.

\section{Appendix I. Variational Notation and Reciprocal Principles}

\vspace{5mm}

``... since $\delta y$ is the variation of $y$ with $x$ kept constant, we could conveniently
write it as $(\delta y)_x$, in accordance with a convention used in thermodynamics...  It is
curious that in spite of the obvious need in partial differentiation for precise statement
of what is being kept constant, such statement is not embodied in customary notation of pure
mathematics, though it is provided in thermodynamics,..."

\vspace {3mm} \noindent
{\it H.Jeffreys and B.S.Jeffreys\\
Methods of Mathematical Physics\\Cambridge U.P.,1946.}

\vspace {6mm}

Variational problems ask for the function which makes stationary
an integral involving the function and its derivatives.
Usually there is a constraint which is fixed. For example,
we are to find a function $y(x)$, describing a closed planar curve,
of fixed length $L$, which maximizes the area $A$ enclosed by the curve
$y(x)$. $A$ and $L$ are functionals of $y(x)$: $A=A[y(x)], L=L[y(x)]$.
For our purpose here, we do not need the precise form
of $A$ or $L$. $L$ is the constraint and $A$ is the functional to be varied.
We use the following notation to describe this variation
$$
(\delta A)_{L} = 0,
\eqno(A1.1)
$$
where the constraint $L$ is indicated as a subscript on the bracket enclosing
the variation $\delta A$ of the functional $A$. The variation $\delta A$ is first-order
 in $\delta y$, and is called the first variation when higher-order variations are considered \cite{Bru} .
 The constraint $L$ is denoted explicitly in (A1.1)
 because it is the most important, or the one we wish to focus on. There may be other
 constraints which are left implicit, e.g. that the planar curve pass through a given point ($x_B, y_B$).
 In this paper we always leave fixed end-position constraints on curves (closed or open) implicit.

The solution of the problem (A1.1) is the function $y(x)$, which satisfies
a differential equation, the Euler-Lagrange equation. The reciprocal
variational problem has the constraint $L$ and the varied functional $A$ interchanged.
In other words, it is described by the equation
$$
(\delta L)_A = 0.
\eqno(A1.2)
$$
The two equations, (A1.1) and (A1.2), \underline{have the same solution} $y(x)$. The reciprocal
variation is searching for the function $y(x)$ which makes stationary (here minimizes) the length $L$ of a
closed curve with fixed enclosed area $A$.
The common solution for both problems is a circle, as is clear intuitively.
One way to prove reciprocity is to use a trick,
due to Lagrange, called the
method of Lagrange multipliers. The problem (A1.1) is first restated in unconstrained form as
$$
\delta A - \lambda \delta L = 0,
\eqno(A1.3)
$$
where the variations of both  $A$ and $L$ are unconstrained and $\lambda$ is an undetermined multiplier.
To obtain (A1.2) from (A1.3), we simply specify that we now fix $A$  (i.e. set $\delta A = 0$) and vary $L$,
i.e.  $(\delta L)_A = 0$. This proves the reciprocity theorem.

Reciprocity is clear intuitively
in the isoperimetric problem (A1.1). An example of reciprocity from physics is the Gibbs conditions
for equilibrium in thermodynamics \cite{Gibbs}. The reciprocal conditions $(\delta S)_U=0$ and
$(\delta U)_S=0$ are both conditions for thermal equilibrium, where $S$ is the entropy and $U$ the
internal energy. These express the facts that the state of thermal equilibrium is that state which
maximizes the entropy for fixed internal energy, and also is that state which minimizes
the internal energy for fixed entropy. (The unconstrained form \cite {109a} is $\delta U - T \delta S = 0$, where the Lagrange multiplier is the temperature $T$. Here the implicit constraints in all cases are fixed system volume $V$ and mole numbers $N_{\alpha}$)
Reciprocal VP's in mechanics have only recently been studied \cite{5,6} - see Sec.2.

This is all we shall need from the
calculus of variations; for more details one should consult a text-book \cite{Ak}.

\section{Appendix II. Equivalence of the Maupertuis and General Maupertuis Principles}

\vspace{6mm}

In Sec. 2.2, for conservative systems, we discussed the equivalence of the Hamilton Principle (HP)
$$
(\delta S)_T = 0,
\eqno(A2.1)
$$
with the General Maupertuis Principle (GMP)
$$
(\delta W)_{\bar E} = 0.
\eqno(A2.2)
$$
Here we show directly the equivalence of the GMP (A2.2) with the original Maupertuis Principle (MP)
$$
(\delta W)_{E} = 0.
\eqno(A2.3)
$$
In all these principles the end-positions $q_A$ and $q_B$ are held fixed.
Since (A2.1), (A2.2) and (A2.3) all generate the same (Euler-Lagrange) equations of motion, it is clear they
must be equivalent, and it is interesting to show directly the equivalence of (A2.2) and (A2.3).
We must show that relaxing the constraint of fixed $E$ in (A2.3) to fixed $\bar E$ in (A2.2)
does not generate spurious or unphysical solutions.

The constraint of fixed $E= H(q, p)$ in (A2.3) applies along and between the trial trajectories. To relax this
constraint,
we use the method of Lagrange multipliers \cite{note4}. Since $H(q, p)= const$, we have $\delta H=0$ , and hence
$$
\lambda \int_0^T \delta H  dt = 0,
$$
where $\lambda$ is an arbitrary constant. The trajectories all start at $t=0$ at fixed point $A$
and end at $t=T$ (which may vary with the trial trajectory) at fixed point $B$. We release the
constraint of fixed $E$ in (A2.3) and choose $\lambda$ such that

$$
\delta W= \lambda\int_0^T \delta H  dt ,
\eqno(A2.4)
$$
where $\delta$ indicates a variation from the true trajectory value and $\delta H \not=0$ in (A2.4). This can
be rewritten as

$$
\delta W = \lambda[\delta\int_0^TH dt  -  H(T)\delta T ],
\eqno(A2.5)
$$

\vspace{2mm} \noindent
where $H(T) \equiv H(q(T), p(T))$, and where we recall that the end-time $T$ is not fixed in (A2.3) .
 Recalling the definition $\bar E=\int_0^T H dt /T$ enables us to write (A2.5) as
 $$
\delta W = \lambda\delta(T \bar E)  - \lambda E \delta T,
\eqno(A2.6)
$$
where we have used the fact that $H=E=const$  \underline{on} the true trajectory. Expanding (A2.6) gives

$$
\delta W = \lambda \bar E  \delta T +  \lambda T  \delta \bar E- \lambda E \delta T.
\eqno(A2.7)
$$
The first and third terms on the right hand side of (A2.7) cancel, since $\bar E= E= const$
\underline{on} the true trajectory,
as assumed in the MP (A2.3).This means
$$
\delta W =   \lambda T  \delta \bar E.
\eqno(A2.8)
$$

To see that $ \lambda =1$ in (A2.8), we apply (A2.8) to the special case of two true trajectories,
with actions $W$ and $W+dW$
 and energies $E$ and $E+dE$. For this case (A2.8) reads $\partial W/ \partial E = \lambda T$.
 Comparison of this with the standard
relation \cite{1b}  $\partial W/ \partial E =  T$ gives $\lambda =1$. Hence we have

$$
\delta W =   T  \delta \bar E,
\eqno(A2.9)
$$
which is the unconstrained Maupertuis Principle (UMP) (2.10). Fixing $\bar E$ in (A2.9) then gives (A2.2), the GMP.

 Thus  (A2.2) follows from  (A2.3). The converse, the derivation of   (A2.3) from  (A2.2) is easy;
 we simply restrict the trial trajectories in
 (A2.2) to those of fixed energy $E$, since the equations
 of motion following from (A2.2) imply energy conservation. Hence
 (A2.2) and   (A2.3) are equivalent.

 We note once more that conservation of energy is assumed in the original MP (A2.3),
 whereas it is a consequence of the general MP (A2.2).


\begin{thebibliography}{99}
\bibitem {0a}  For example, in Yourgrau and Mandelstam \cite{2} the notation is reversed. Hamilton himself used the current notation, but, for the actions
 along true paths, termed $S$ and $W$ the principal and characteristic functions, respectively.
\bibitem {14a} Holonomic (coordinate) constraints are assumed to be taken into account by the choice of the $q_1, q_2, ..., q_f$. Nonholonomic (velocity) constraints (see note \cite{Non}) are assumed absent in the early sections but are considered briefly in sec.6 and App.II.
\bibitem{Cas} The condition $H=K+V$ is valid for normal systems, for which $L=K-V$, the holonomic constraints (if any) are time-independent and the reference frame is inertial. If these conditions are satisfied $K$ is quadratic in the $\dot {q}$'s. Note that the single condition $K$ quadratic in the $\dot {q}$'s is not sufficient to have $H=K+V$ since for some systems with time-dependent constraints $K$ is quadratic in the 
$\dot {q}$'s: see, e.g. de Castro A S 2000 {\it Eur. J. Phys.} {\bf 21} 23 ; 
Ferrario C and Passerini A 2002 {\it Eur. J. Phys.} {\bf 22} 11.

When $H\not= K+V$, the generalized Maupertuis principles require $\bar{H}$ in place of $\bar E$, 
where $H = p\dot{q}-L$ (see Secs. 5.2.1 and 6).
\bibitem {24a} This particular analogy can be pushed further. Just as there are numerous other
thermodynamic potentials besides the Helmholtz free energy $F$ and the Gibbs free energy 
$G = F + pV$, (where $p$ is the pressure and $V$ the volume), e.g. the  internal energy 
$U = F + TS$, (where $T$ is the temperature and $S$ the entropy), we can introduce in mechanics further
 actions via further Legendre transformations. For example, define $S^{'}=S - p_B q_B$,
 where  $q_B$ is the final end-point position and
 $p_B$ the true momentum at $B$. If we relax the constraint of fixed $q_B$, the UHP generalizes further 
to $\delta S = -E \delta T + p_B \delta q_B$. In terms of $S^{'}$ this transforms to 
$\delta S^{'} = -E\delta T - q_B \delta p_B$. In terms of $S^{'}$, the usual fully constrained HP is 
$(\delta S^{'})_{T, p_B} = 0$, replacing $(\delta S)_{T, q_B} = 0$. Here, as always, the constraint on $q_A$ is still  
left  implicit in both. Further Legendre (using $p_A, q_A$ and/or $W$)
and reciprocal transformations are obvious.
In this review we focus on the four basic variational principles summarized at the end of sec.2.2.
\bibitem {35a} The RMP method yields $\omega_{x}/\omega_{y}= W_{y}/W_{x}$ 
and hence after 
semiclassical quantization $\omega_{x}/\omega_{y} = (n_{y}+\frac{1}{2})/(n_{x}+ \frac{1}{2})$ , the same as (4.7) 
found by the fully quantum VP. Since 
$\omega_{x}/\omega_{y}$ is therefore rational, the optimum quantized trajectories are periodic, at least at this 
level of approximation. Periodic trajectories also play a key role in Gutzwiller's trace formula method of semiclassical quantization \cite{9a}.
\bibitem {42f}\underline{ Exohedral} complexes XC$_n$, where X is on the \underline {outside} of the cage, 
have also been studied, e.g. H$_2$C$_{60}$ \cite {42g}
 and H$_2$C$_{\infty}$ (nanotubes) \cite{42h}. Complexes such as K$_3$ C$_{60}$ are superconductors in the solid  state \cite {42b}.
\bibitem {42pa} The potential $V(r)$ can be fit \cite{42d} to simple analytic forms(polynomial, Morse, etc). In some cases (e.g. Li$^{+}$@C$_{60}$) the potential has its minimum off-centre. 
For K$^{+}$@C$_{60}$ the potential is essentially $V(r) = C r^4$, a rare example of a pure quartic oscillator in nature.
\bibitem  {note3} The form $ \delta W = \int_0^T \delta H dt$ for the UMP, which is easily shown to be
equivalent to (5.4), essentially
appears in passing in the equations of Yourgrau and Mandelstam \cite{2},p.46, but they do not note explicitly its
validity for nonconservative systems. For the special case of periodic systems evolving adiabatically
with parameter $\lambda(t)$, Ehrenfest \cite{Ehr} gives a relation related to (5.4); he has $\delta E$ rather
than $\delta \bar E$   (as in the original MP),  and our $\delta T$  is replaced by a $d\lambda$ term. Ehrenfest attributes his
 result to Boltzmann and Clausius. 

As before, in the UMP (5.4) we have fixed the initial time
$t_A = 0$ and allow the final time $t_B \equiv T$ to vary. If both $t_A$ and $t_B$ 
are allowed to vary, (5.4) generalizes to $T\delta \bar E - \delta W - [ E(t_B)- \bar E] \delta t_B + [ E(t_A)- \bar E] \delta t_A = 0$. This reduces to (5.4) for fixed $t_A$, and to (2.10) 
for conservative systems ($E(t_A) = E(t_B) = \bar E$). Further generalizations, allowing $q_B$ and/or $q_A$ to vary, are discussed in notes 
\cite {note2, 24a}
\bibitem {note7} The  converse can be made plausible: from an assumed adiabatic invariance of the
long path $W$, the adiabatic invariance of $\Omega$ follows. We use \cite{Tab}, for $T\to\infty$,
$W\to \oint _C\sum_i p_i{dq_i}=\sum_i\int\int_{R_i} {dq_i}{dp_i}$, where $R_i$ is the projected area
of curve $C$ on the $q_i, p_i$ plane,  and the fact that
the dynamical invariance (Poincar\'{e} invariant) of  $\Omega=\int...\int\ {dq_1}{dp_1}...{dq_f}{dp_f}  $
can be derived \cite{Tab} from the dynamical invariance of $ \sum_i \int\int{dq_i}{dp_i},$

The adiabatic invariance of long-path  $W$'s for ergodic systems follows heuristically from the UMP (5.4)
as described in the text. Alternatively, we can heuristically argue that for $T \to \infty$, the long path is essentially
closed in phase space, so that the Poincar\'{e} dynamical invariant $W\equiv \oint\sum_i p_i{dq_i}$
remains essentially adiabatically constant as the energy
surface and the path on it slowly evolve. Obviously the limiting procedures are delicate and problematic.
Even if our conjecture is true, the adiabatic invariance may appear to be of the seemingly uninformative type $\infty=\infty$,
since $W\to\infty$ for $T\to\infty$. To extract finite results, $T$ will need to be taken as, say, the Poincar\'{e}
cycle time. We note that, in another situation where $W\to\infty$ as
$T\to\infty$, i.e. scattering orbits, useful results can still  be extracted \cite{5}.

The arguments in the above paragraph, as well as the one in the text, apply also more generally to the chaotic motions of 
nonergodic nonintegrable systems. Since adiabatic invariants have not been found previously for such systems, it would be of interest to 
prove, or disprove, these suggestions. If proven, it would then be of interest to establish a general semiclassical 
quantization scheme for nonintegrable systems based on $W$ (generalizing (5.32) which holds for ergodic systems).
\bibitem {52} The cumulative number of states up to $E$,  $N(E)$,
 is defined quantum mechanically by  $N(E)=\sum_n \Theta(E-E_n)= {\rm tr}\;  \Theta( E- \hat H)$, where $\hat H$ is the
 Hamiltonian operator and tr denotes a trace.
\bibitem {Non} Nonholonomic constraints $f_i(q, \dot q, t)=0$  contain the velocities $\dot q_{\alpha}$, such that the
 constraints cannot be integrated up to global coordinate  constraints $g_i(q, t)=0$.\underline{ Kinematic}
 nonholonomic constraints arise from
 introducing redundant variables as in sec.6 on relativistic mechanics, or by introducing constants
 of the motion as constraints \cite{Kap} as in App.II for the MP, or by imposing the
 kinematic constraint between $\dot q$ and
 the velocity or the momentum after first having released the constraint \cite{Leub}.
 \underline{ Dynamic} nonholonomic constraints arise
 from physical forces of constraint, such as arise in rolling motion \cite{16a}.
 Kinematic nonholonomic constraints can be handled by an effective
 Lagrangian $L_{eff} = L + \sum_i \lambda_i f_i$,
 where $\lambda_i$ are Lagrange multipliers \cite{Lan, Bar,For}. Dynamic nonholonomic constraints
  cannot be handled by such an effective Lagrangian (contrary
 to what is stated in some excellent textbooks) and construction of an HP or MP is highly problematic \cite {Pap}.
\bibitem{16a}
Two classic rolling motion examples, both of  which are integrable
systems, are the nonslipping tippe top \cite{16b} and the
nonslipping Euler disc (or spinning coin) \cite{16c}.
\bibitem {106a} Another common choice is ${\bf A} = ( 0, xB, 0)$. This choice is known as the Landau gauge (in this gauge {\bf A} is invariant under translations along the y axis). The first choice is known as the symmetric gauge (because of the rotational invariance of {\bf A} about the z axis).
\bibitem {81a} There are various ways to constrain the variation so that $\psi_n$ is selected.
In principle, one can minimize the energy $<\psi| \hat {H}|\psi>/<\psi| \psi> $ 
with the trial $\psi$'s restricted to be 
orthogonal to the exact $\psi_0, \psi_1, ..., \psi_{n-1}$. In practice, since $\psi_0, \psi_1, ..., \psi_{n-1}$ are rarely known exactly, one can find an approximate $\psi_n$ by choosing trial $\psi$'s
 with $n$ nodes (for 1D problems). Alternatively, one can choose a trial $\psi$
 which is known to be the $n$th eigenstate of a similar Hamiltonian (see sec. 3.1 for an example).
\bibitem{95a} In practice (7.18) is less useful for bound states than (7.1) since $E$ and $\psi$ are both unknown for bound states. For scattering states $E$ is known, and  $<\psi|\psi>\to \infty$ causes no problem in (7.18), as it does in (7.1).
\bibitem {z1} The derivation of the Einstein gravitational field equations from the Einstein - Hilbert action is the classic example:
 see, e.g., Landau and Lifshitz \cite {14}, p 307
\bibitem {109a} This is usually written as $\delta F =0$, where $F=U-TS$ is the Helmholtz free energy: i.e. the free energy is a minimum at 
equilibrium at a given $T$, $V$ and $N_{\alpha}$.
\end{thebibliography}
\end{document}